\documentclass[%
 reprint,
 amsmath,amssymb,
 aps,
]{revtex4-2}

\usepackage{physics}
\usepackage{graphicx}% Include figure files
\usepackage{dcolumn}% Align table columns on decimal point
\usepackage{bm}% bold math
\usepackage{array}
\begin{document}

\preprint{APS/123-QED}

\title{Exchange Monte Carlo for continuous-space Path Integral Monte Carlo simulation}% Force line breaks with \\

\author{Xun Zhao}
\affiliation{Department of Physics, The University of Tokyo, 7-3-1 Hongo, Bunkyo-ku, Tokyo 113-0033, Japan}
\author{Synge Todo}%
\affiliation{Department of Physics, The University of Tokyo, 7-3-1 Hongo, Bunkyo-ku, Tokyo 113-0033, Japan}
\affiliation{Institute for Physics of Intelligence, The University of Tokyo, 7-3-1 Hongo, Bunkyo-ku, Tokyo 113-0033, Japan}
\affiliation{Institute for Solid State Physics, The University of Tokyo, 5-1-5 Kashiwanoha, Kashiwa, Chiba 277-8581, Japan}

\date{\today}% It is always \today, today,

\begin{abstract}
We present a novel Exchange Monte Carlo (EMC) method designed for application in continuous-space Path Integral Monte Carlo (PIMC) simulations at finite temperature. Traditional PIMC methods for bosonic systems suffer from long autocorrelation times, particularly when measuring observables affected by particle permutations, such as the winding number. To address this issue, we introduce an exchange update scheme that facilitates replica transitions between different interaction regimes, significantly accelerating Monte Carlo dynamics—especially for global observables sensitive to permutation effects. Furthermore, we incorporate Stochastic Potential Switching (SPS) to efficiently decompose interactions, substantially enhancing computational efficiency for long-range interatomic pair potentials such as the Lennard-Jones and Aziz potentials. 
\end{abstract}

\maketitle

%\tableofcontents

\section{\label{sec:level1}INTRODUCTION}

Markov Chain Monte Carlo (MCMC) methods have been among the most important computational techniques across various fields since their inception~\cite{landau2021guide,liu2001monte,brooks1998markov, eraker2001mcmc}. The fundamental principle of MCMC methods is to generate a sequence of random configurations to approximate high-dimensional integrals efficiently. In statistical physics, Path Integral Monte Carlo (PIMC) remains the only known method capable of simulating bosonic systems and providing numerical estimates of physical quantities at finite temperatures. The primary challenge in these simulations is efficiently incorporating exchange interactions for large system sizes, as local updates struggle to sample different worldline topologies. The conventional Monte Carlo method proposed by Ceperley and Pollock~\cite{pollock1987path, sindzingre1989path} was the first to yield quantitative results for superfluid systems; however, it encounters difficulties in sampling long permutation cycles as system size increases. No significant breakthroughs were made until the development of the worm algorithm by Boninsegni, Prokof’ev, and Svistunov~\cite{boninsegni2006worm}, which enabled simulations of superfluid $^4$He with hundreds of particles in the grand canonical ensemble, yielding results consistent with experimental data.  

Despite its broad success, the worm algorithm is most naturally formulated in the grand-canonical
ensemble and typically requires careful tuning of several proposal parameters to maintain
reasonable acceptance rates. A representative example is the choice of how many beads
(imaginary-time slices) to modify in a single worm update. On the one hand, proposing a longer
segment increases the chance that the worm head reconnects with the tail and thereby changes the
worldline connectivity. On the other hand, as the updated segment spans more time slices, the
proposal becomes increasingly likely to be rejected due to strong short-range repulsion with beads
at the same imaginary-time slices. In practice, this results in long autocorrelation times for
topological quantities: the Markov chain can remain trapped in a fixed winding-number sector for
extended periods, substantially reducing sampling efficiency.

This dynamical trapping is reminiscent of frustrated systems, where exchange Monte Carlo
(EMC) \cite{hukushima1996exchange} is routinely used to accelerate equilibration by allowing replicas at different conditions to
swap configurations. In its standard form, EMC employs a temperature ladder, leveraging the faster
local dynamics at high temperature to facilitate barrier crossing at low temperature. However, in
PIMC, high-temperature replicas strongly suppress long permutation cycles and
therefore do not effectively promote transitions between winding-number sectors that dominate the
low-temperature physics. As a result, conventional temperature-based EMC provides limited
improvement to the mixing of winding-number sectors in the regime of interest.

In this work, we present an exchange-based strategy tailored to canonical-ensemble PIMC that
directly targets this bottleneck. In Sec.~II, we review the standard PIMC formulation and the
canonical-ensemble implementation of worm-type updates used to sample bosonic exchange. We then
describe the local update components of our approach, including the No-U-Turn Sampler (NUTS)~\cite{hoffman2014no}
and stochastic potential switching (SPS)~\cite{mak2005stochastic}, which together improve sampling
efficiency and reduce computational cost. The central contribution is the introduction of auxiliary
replicas and exchange moves within the PIMC framework to accelerate the dynamics of the physical
replica, thereby enhancing transitions between winding-number sectors. In Sec.~III, we summarize
estimators for physical observables in the presence of SPS and a nonuniform discretization of
imaginary time. In Sec.~IV, we present numerical results for superfluid \(^4\)He that demonstrate the
benefits of the proposed exchange updates. Finally, in Sec.~V, we discuss possible extensions and
conclude the paper.

\section{ALGORITHMS}
\subsection{PIMC framework and Worm update}
We start from the quantum canonical partition function
\begin{equation}
    Z = \Tr\!\left[e^{-\beta \hat{H}}\right],
\end{equation}
where \(\beta\) is the inverse temperature. For a system of \(N\) identical particles in \(D\) spatial dimensions, we denote the many-body configuration and momentum operators by
\begin{align}
    \hat{\mathbf{x}} &\equiv (\hat{\mathbf{x}}_1,\ldots,\hat{\mathbf{x}}_N), \\
    \hat{\mathbf{p}} &\equiv (\hat{\mathbf{p}}_1,\ldots,\hat{\mathbf{p}}_N)
\end{align}
with \(\hat{\mathbf{x}}_a\) and \(\hat{\mathbf{p}}_a\) the position and momentum operators of particle \(a\), respectively.
The Hamiltonian is written as
\begin{equation}
    \hat{H} = \sum_{a=1}^{N}\frac{\hat{\mathbf{p}}_a^{\,2}}{2m} + V(\hat{\mathbf{x}}),
\end{equation}
where \(V(\hat{\mathbf{x}})\) is an arbitrary many-body potential that depends on the full configuration \(\hat{\mathbf{x}}\) (e.g., external and/or interaction terms).

Using the eigenbasis $\ket{\mathbf{x}}$ of $\hat{\mathbf{x}}$, where $\mathbf{x}\in\mathbb{R}^{ND}$, the trace becomes an integral over the full configuration space:
\begin{equation}
    \begin{split}
        Z 
        &= \int d\mathbf{x}\, \mel{\mathbf{x}}{e^{-\beta \hat{H}}}{\mathbf{x}}
         = \int d\mathbf{x}\, \mel{\mathbf{x}}{\left(e^{-\tau \hat{H}}\right)^P}{\mathbf{x}},
    \end{split}
\end{equation}
where $P$ is the number of Trotter slices and $\tau \equiv \beta/P$ is the imaginary-time step.
Inserting the resolution of identity $P-1$ times,
$\int d\mathbf{x}_j\,\ket{\mathbf{x}_j}\bra{\mathbf{x}_j}=\hat{I}$ (\(j=2,\ldots,P\)),
regarding \(\mathbf{x}\) as \(\mathbf{x}_1\), and applying the Suzuki--Trotter decomposition~\cite{suzuki1976generalized, trotter1959product}, we obtain the standard discretized path-integral form
\begin{align}
    Z &= (4\pi\lambda\tau)^{-\frac{PND}{2}}
    \int \prod_{j=1}^{P} d\mathbf{x}_j\,
    \exp\!\left[-S_P(\{\mathbf{x}_j\})\right],
    \label{eq:partition}
\end{align}
with
\begin{align}
    S_P(\{\mathbf{x}_j\})
    &= \sum_{j=1}^{P}\left[
    \frac{\lVert \mathbf{x}_{j+1}-\mathbf{x}_j\rVert^2}{4\lambda\tau}
    + \tau\,V(\mathbf{x}_j)\right], \\
    \mathbf{x}_{P+1} &\equiv \mathbf{x}_1.
\end{align}

where $\mathbf{x}_j\in\mathbb{R}^{ND}$ is the $ND$-component configuration at Trotter slice $j$ (the ``beads''), $\|\cdot\|$ is the Euclidean norm on $\mathbb{R}^{ND}$, and $\mathbf{x}_{P+1}\equiv \mathbf{x}_1$ enforces the $\beta$-periodic (closed) boundary condition in imaginary time.
Here and in the following, we set the reduced Planck constant $\hbar=1$ and the mass of the particles $m=1$, so that the kinetic-energy scale is $\lambda \equiv \hbar^2/(2m) = 1/2$. Energies are measured in units where $k_{\mathrm B}=1$.

For indistinguishable particles, the path-integral configurations must also include particle permutations: worldlines can reconnect such that \(\mathbf{x}_{j,a}\) at one slice may be connected to a different particle label at another slice. In practice, to sample these permutation sectors efficiently within PIMC, we incorporate the \emph{open}, \emph{close}, and \emph{swap} updates of the worm algorithm (or their canonical-ensemble variants)~\cite{boninsegni2006worm, spada2022path}, which allow the Markov chain to change the permutation (worldline connectivity) in addition to the bead positions.

\subsubsection{Open and Close update}
Directly swapping two paths typically requires repositioning many beads, which can result in a very low acceptance rate. Consequently, it is advantageous to decompose a swap update into three sequential steps: (i) negating one spring interaction (make a cut) along the path (open update), (ii) connecting one end of the cut to a different particle index (swap update), and (iii) reconnecting the cut (close update). This approach maintains a higher acceptance probability. Moreover, exploring configurations with a single cut, which is called the G sector in convention, is a natural strategy for measuring the condensate fraction of the system~\cite{boninsegni2006worm}.

The details of the open update procedure are as follows:
\begin{enumerate}
    \item A particle \( a \) is randomly selected from the \( N \) particles, and a slice \( j \) is randomly chosen from the \( P \) slices.
    \item All beads from slice \( j+1 \) to slice \( j+m-1 \) are removed, where \( m \) is a hyperparameter that controls the length along the imaginary time axis.
    \item A new path is constructed from slice \( j+1 \) to slice \( j+m \) using the Lévy construction~\cite{bertoin1996levy}, thereby creating a cut between slice $j$ and slice $j+1$.
\end{enumerate}
This procedure transitions the simulation into the G sector by removing the interaction between slices \( j \) and \( j+1 \) for particle \( a \). The acceptance probability, following the general Metropolis-Hastings rule~\cite{metropolis1953, hastings1970monte}, is 
\begin{equation}
    A_\text{op} = \min\Big[ 1, C_1 \exp(-\tau \Delta V)\exp{\frac{\|\mathbf{x}_{j+m,a} - \mathbf{x}_{j,a}\|^2}{2\tau m}}\Big],
    \label{open update}
\end{equation}
where
$C_1 = N P \left(2\pi \tau m\right)^{D/2} C_1'$
and \( C_1' \) is a hyperparameter controlling the ratio between configurations in the Z sector (i.e., configurations without a cut) and those in the G sector. Here, \(\Delta V\) denotes the difference in interacting potential energy and external potential energy between the old path and the newly generated path.

The close update complements the open update, enabling transitions from the G sector back to the Z sector. The procedure for the close update is as follows:
\begin{enumerate}
    \item Identify the location of the cut in the configuration (e.g., at particle \( a \) and slice \( j \)).
    \item Remove all beads from slice \( j+1 \) to slice \( j+m-1 \).
    \item Construct a new path from slice \( j \) to slice \( j+m \) using the Lévy construction, thereby producing a fully connected path in the Z sector.
\end{enumerate}
The acceptance probability for the close update is given by
\begin{equation}
    A_\text{cl} = \min\Big[ 1, \frac{1}{C_1} \exp(-\tau \Delta V)\exp{-\frac{\|\mathbf{x}_{j+m,a} - \mathbf{x}_{j,a}\|^2}{2\tau m}}\Big].
\end{equation}

\subsubsection{Swap update}
The swap update is a self-complementary procedure that modifies the path permutation. It is only proposed when the configuration is in the G sector (i.e., when a cut is present along the path). The swap update proceeds as follows:
\begin{enumerate}
    \item Identify the position of the cut, for example, at particle \( a \) and slice \( j \).
    \item Compute the total weight \( W_{\text{old}} \) of the path connecting the current bead at \( (j, a) \) to the bead at slice \( j+m \), defined as
    \begin{equation}
        W_{\text{old}} = \sum_{c=1}^{N} \exp{-\frac{\|\mathbf{x}_{j+m,c}-\mathbf{x}_{j,a} \|^2}{2\tau m}}.
    \end{equation}
    \item Use tower sampling (or any other discrete variable sampling technique) to select a particle \( b \) according to its weight.
    \item Compute the total weight \( W_{\text{new}} \) associated with \( b \) in the same manner as \( W_{\text{old}} \):
    \begin{equation}
        W_{\text{new}} = \sum_{c=1}^{N} \exp{-\frac{\|\mathbf{x}_{j+m,c}-\mathbf{x}_{j,b} \|^2}{2\tau m}}.
    \end{equation}
    \item Construct a Lévy path from particle \( a \) at slice \( j \) to particle \( b \) at slice \( j+m \).
\end{enumerate}
The acceptance probability for this proposal is given by
\begin{equation}
    A_{\text{swap}} = \min\left\{ 1,\, \exp\left(-\tau \Delta V\right) \frac{W_{\text{old}}}{W_{\text{new}}} \right\}.
\end{equation}

\subsection{No U-turn Sampler (NUTS) and Stochastic Potential Switching (SPS)}

For the local update of bead positions, we adopt Hamiltonian Monte Carlo (HMC), which is generally considered superior to traditional random-walk MCMC methods in continuous-space models~\cite{duane1987hybrid, betancourt2017conceptual, neal2011mcmc}. In HMC, a next state is proposed by using the virtual Hamiltonian dynamics. A primary drawback of HMC is the difficulty in determining an appropriate stopping criterion for the Hamiltonian evolution. The No-U-Turn Sampler (NUTS) addresses this challenge by automatically selecting the number of integration steps by enforcing a no-U-turn condition~\cite{hoffman2014no}, thereby eliminating the need for manual tuning of update parameters.

Furthermore, we apply stochastic potential switching (SPS)~\cite{mak2005stochastic} to treat pairwise interactions with an $O(N)$ computational cost. 
As a benchmark for \(^4\)He, we employ the Aziz potential~\cite{aziz1979accurate}, which is widely used as a realistic interatomic interaction.

For a pair potential $U(r)$ that depends only on the interparticle distance $r\equiv|\mathbf{x}_a-\mathbf{x}_b|$, SPS stochastically replaces $U(r)$ by either a ``switched'' potential $\tilde U(r)$ with probability $S(r)$ or a complementary potential $\bar U(r)$ with probability $1-S(r)$ at each update.
Following Ref.~\cite{mak2005stochastic}, we introduce the switching probability $S(r)$ as
\begin{equation}
S(r) = e^{\tau [U(r) - \tilde U(r) - \Delta U^{*}]},
\end{equation}
with a constant$\Delta U^{*} \ge \max_r [U(r) - \tilde U(r)]$ that guarantees $S(r)\in[0,1]$ for all $r$.
The complementary potential $\bar U(r)$ is defined as
\begin{equation}
\bar U(r)= U(r) - \frac{1}{\tau} \ln [ 1-S(r) ],
\end{equation}
so that the identity
\begin{equation}
e^{-\tau U(r)} = e^{-\tau [\tilde U(r) + \Delta U^{*}]} + e^{-\tau \bar U(r)},
\end{equation}
is satisfied. 
As we enforce the detailed balance in the dynamics generated under $\tilde U(r)$ and under $\bar U(r)$, the combined SPS procedure satisfies detailed balance for the original potential $U(r)$~\cite{mak2005stochastic}.
In practice, the choice of $\Delta U^{*}$ affects efficiency but not correctness; when $U(r)-\tilde U(r) \le 0$ for all $r$, one may simply set $\Delta U^{*}=0$.

The remaining question is how to choose $\tilde U(r)$ to accelerate the simulation.
Here we adopt a Weeks--Chandler--Andersen (WCA)-type decomposition~\cite{weeks1971role}:
\begin{equation}
\tilde U(r)=
\begin{cases}
U(r)+\epsilon, & r \le r_c,\\
0, & r > r_c,
\end{cases}
\end{equation}
where $r_c$ is the distance at which $U(r)$ attains its minimum and $\epsilon\equiv |U(r_c)|$ is the depth of that minimum.
For typical Lennard--Jones-type interactions, the attractive tail decays rapidly with $r$, so $|U(r) - \tilde U(r)|$ becomes small at large separations, and the switching failure probability $1-S(r)$ is correspondingly small for $r>r_c$.
Consequently, the expected number of ``failed'' pairs beyond $r_c$ is $O(1)$ per particle at fixed number density, yielding an overall $O(N)$ cost.

Keeping track of particles within a distance $r_c$ can be easily achieved by the binning technique common in molecular dynamics~\cite{hollingsworth2018molecular}. We treat the switching of potential at a distance larger than $r_c$ by two complementary updates. 

\begin{figure}[tbp]
    \centering
    \includegraphics[width=0.4\textwidth]{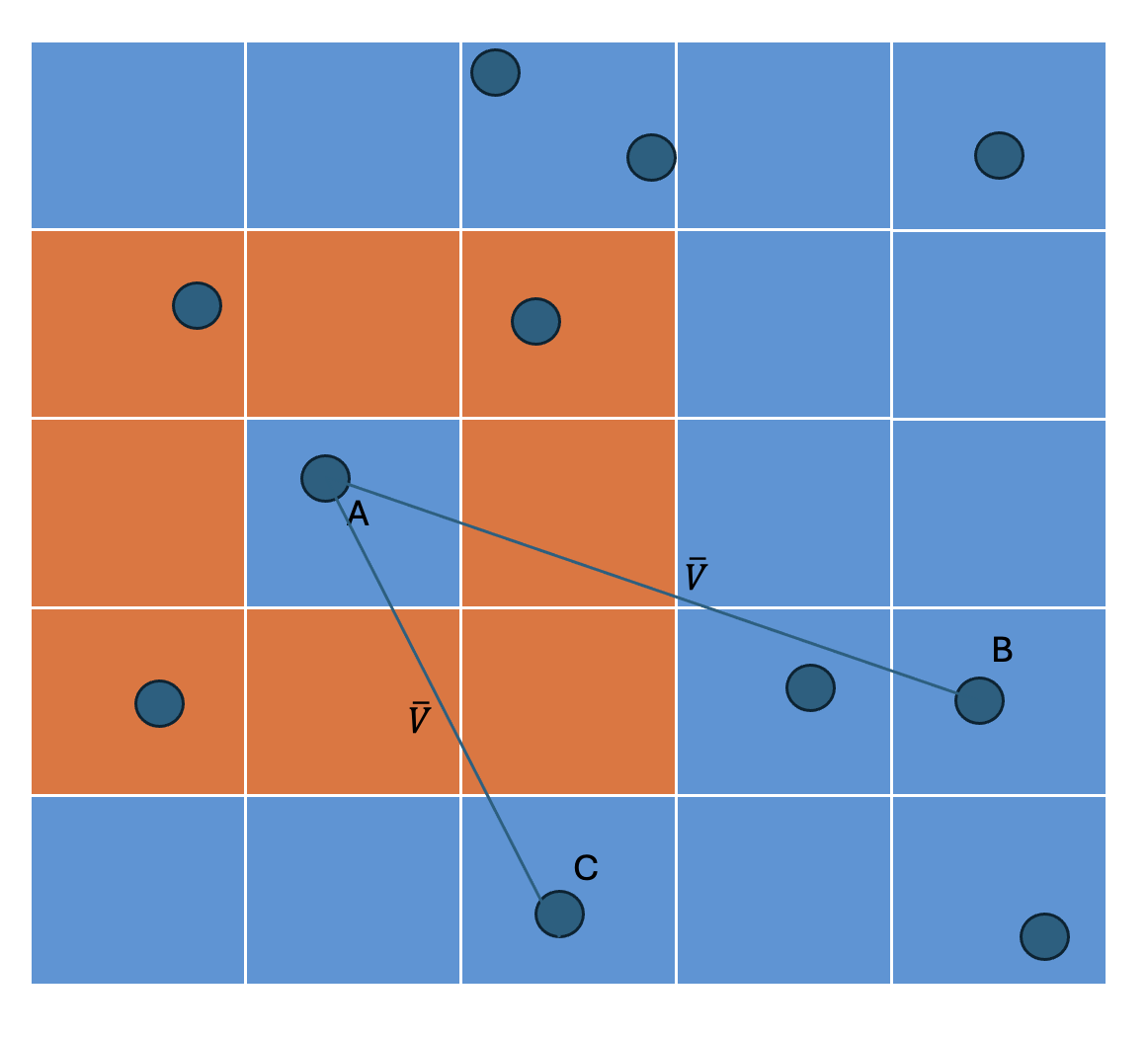}
    \caption{Stochastic potential switching treatment of particle interactions in two dimensions. The space is divided into bins of size $r_c$. For interactions of particle A, we only need to consider the nearest bins (orange region) and interactions that are switched to $\bar U$ (particle B and C).}
    \label{fig:sps}
\end{figure}

1. To switch the potential from $\tilde{U}$ to $\bar{U}$ (where $\bar{U}$ is nonzero for $r > r_c$), the update proceeds as follows. First, a particle $a$ is randomly selected from the $N$ particles, and its corresponding bin, labeled $A$, is identified. Next, another bin, $B$, is chosen according to a probability distribution $p_{AB}$ using Walker's method of aliases~\cite{walker1974new}. If bin $B$ contains no particles, the update is rejected. Otherwise, a particle $b$ is randomly selected from bin $B$. If the interaction between $a$ and $b$ is already $\bar{U}$, the update is also rejected. Otherwise, the acceptance probability for the proposed update is given by:
\begin{equation}
    P = \min\left\{ 1, 
    \begin{aligned}
        &\frac{n_A \times n_B}{n_A + n_B} \times 
        \frac{l_a+l_b+2}{(l_a+1)(l_b+1)} \times \\
        &\frac{\exp(-\tau \bar{U}(r))}{\exp(-\tau \tilde{U}(r))} \times 
        \frac{1}{p_{AB}} ,
    \end{aligned} \right\}
\end{equation}
where $n_A$ and $n_B$ denote the number of particles in bins $A$ and $B$, respectively, while $l_a$ and $l_b$ represent the number of interactions assigned $\bar{U}$ for particles $a$ and $b$. 

\vspace{5mm}

2. To switch the potential from $\bar{U}$ back to $\tilde{U}$, the update randomly selects a particle $a$ and identifies all particles interacting with it via $\bar{U}$. If no such particles exist, the update is rejected. Otherwise, a particle $b$ is randomly drawn from this list, and the acceptance probability is given by:
\begin{equation}
    P = \min\left\{ 1, \frac{n_A+n_B}{n_A \times n_B} \times \frac{l_a \times l_b}{l_a+l_b} \times \frac{\exp(-\tau \tilde{U}(r))}{\exp(-\tau \bar{U}(r))} \times p_{AB} \right\}.
\end{equation}
Here, the probability $p_{AB}$ is chosen as:
\begin{equation}
    p_{AB} \sim \frac{\exp(-\tau \bar{U}(r_{AB}))}{\exp(-\tau \tilde{U}(r_{AB}))},
\end{equation}
where $r_{AB}$ is the distance between the centers of bins $A$ and $B$. This choice ensures that the acceptance probability remains of order 1, improving the efficiency of the update.

There are two main advantages of using SPS with WCA decomposition over the diagrammatic Monte Carlo treatment~\cite{boninsegni2006worm}. First, the WCA-decomposed potential function is smooth and free of cusps, making it fully compatible with the HMC/NUTS algorithm. Since calculating forces in HMC requires a smooth potential, this decomposition ensures that updates do not violate detailed balance. Second, in the decomposition equivalent to diagrammatic Monte Carlo, the switched potential $\bar{U}$ diverges within the cutoff distance $r_c$. As a result, position updates can be applied only to particles without active bonds, limiting the algorithm's flexibility. While it has been suggested that careful tuning of $r_c$ can mitigate this limitation, the WCA-based approach eliminates the need for such tuning. With WCA decomposition, all particles can move freely under Hamiltonian dynamics without concerns about potential divergence or cusps, leading to a more robust and generalizable algorithm.

\subsection{Exchange Monte Carlo (EMC)}
\subsubsection{Classical Exchange Monte Carlo}
Exchange Monte Carlo(EMC) is a Monte Carlo simulation method originally developed to explore configuration space more efficiently in systems with a rugged energy landscape, such as spin glass. Such frustrated systems typically exhibit many local minima, making random-walk MCMC difficult to escape. The Exchange Monte Carlo method was originally devised by Swendsen and Wang~\cite{swendsen1986replica} and later extended by Hukushima and Nemoto~\cite{hukushima1996exchange} to overcome energy barriers in Monte Carlo simulations. It was further extended to molecular dynamics simulations~\cite{hansmann1997parallel}.

The core idea of the EMC algorithm is to leverage the dynamics of Monte Carlo updates at high temperatures to help low-temperature configurations escape local minima. A typical exchange update involves two configurations (or replicas), each sampled at a different temperature. The algorithm then attempts to swap their temperatures, with the exchange being accepted or rejected based on the standard Metropolis-Hastings criterion
\begin{equation}
    p = \min(1 , \frac{e^{-\beta_{1}E_{2}}e^{-\beta_{2}E_{1}}}{e^{-\beta_{1}E_{1}}e^{-\beta_{2}E_{2}}}) = \min(1, e^{(\beta_{1}-\beta_{2})(E_{1} - E_{2})}).
\end{equation}
It is straightforward to verify that the detailed balance condition for the global system is satisfied, and this update is independent of any configuration changes at each temperature. Ideally, a replica starting at low temperature will be heated to high temperature, undergo some Monte Carlo updates with fast relaxation at high temperature, and finally return to low temperature to escape from a local minimum. 

Therefore, the efficiency of EMC depends heavily on how often the replica reaches the high- and low-temperature limits. Many attempts have been made to optimize the distribution of replicas in the temperature space, which can be classified mainly into two categories: 1. uniform exchange rate between all replicas~\cite{hukushima1999domain}, 2. minimization of round-trips between temperature limits~\cite{katzgraber2006feedback}. In this paper, we make the first attempt to flatten the exchange rate distribution. Within the PIMC framework, we optimize the distribution of imaginary-time slices rather than the temperature distribution of replicas.

The present construction can also be viewed as a PIMC-specific realization of generalized-ensemble and Hamiltonian replica-exchange methods. In multidimensional replica exchange, the exchanged parameters are not restricted to temperature but may also include parameters in the potential energy \cite{Sugita2000MultidimensionalREM}. Similarly, in Hamiltonian replica exchange, auxiliary replicas are generated by modifying the Hamiltonian so as to enhance configurational sampling \cite{Fukunishi2002HamiltonianREM}. The essential difference in the present work is the choice of the exchange coordinate. Rather than introducing a generic coupling-constant interpolation, we vary the number and distribution of imaginary-time slices at which the interparticle interaction is active. This choice is motivated by the sampling bottleneck specific to bosonic PIMC: in auxiliary replicas with more noninteracting imaginary-time slices, world-line rearrangements are less constrained by interparticle interactions, and updates that change permutation cycles and winding-number sectors can occur more readily. Exchange moves then transfer this improved sampling of global topological sectors back to the fully interacting target replica while keeping the physical temperature fixed.

\subsubsection{EMC in PIMC}
In continuous-space Path Integral Monte Carlo simulations, the main challenge we face is measuring observables related to the topology of ring polymers, which arises from the indistinguishable nature of bosonic particles. For example, the superfluid density of the system is measured by calculating the winding number, a global quantity determined by the system's topology. Local updates, such as a random walk proposal for each particle separately, will not change the configuration's winding number within the periodic box. The key is to first efficiently sample configurations with large permutation cycles, then move the entire polymer together to wind it around the box. Previous methods, such as the bisection method~\cite{ceperley1995path} or the worm algorithm~\cite{boninsegni2006worm}, effectively addressed the ergodicity problem, but efficiency remains a challenge. The reason is that there is always a trade-off when updating the path. In the worm algorithm, for example, we always hope that the starting point and ending point of an updating worm are separated along imaginary time slices as much as possible. In this way, the worm has enough time to evolve and wind around the box within the imaginary time. However, the acceptance rate of the update will decrease as we increase the number of beads along the path, due to the strong bosonic repulsion within each imaginary-time slice. As a result, the configuration may remain in the same winding-number sector for a long time. 

\begin{figure}[tbp]
    \centering
    \includegraphics[width=0.5\textwidth]{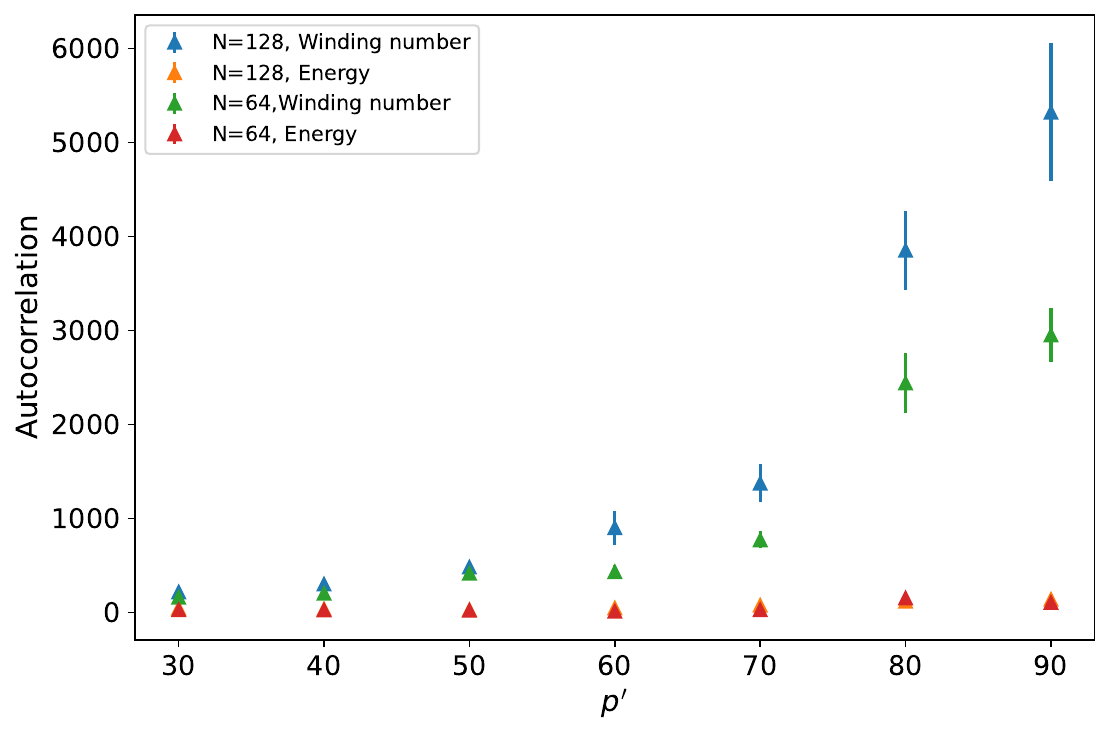}
    \caption{Autocorrelation time of observables vs. number of interacting imaginary slices $p'$. Autocorrelation time quantifies how long it takes for an observable in a time series to become statistically independent, and it is calculated using binning. Superfluid fraction, which is affected by the configuration's topology, can be sampled more efficiently during simulations for partially interacting systems. }
    \label{fig:atuo_vs_inter_slice}
\end{figure}

To take advantage of the exchange update, we need to prepare replicas in a regime where the dynamics of the Monte Carlo update are much faster. Instead of a high temperature limit, which is not helpful in terms of sampling large permutation cycles in PIMC, we notice that it is much easier to explore configurations with large permutation cycles for free bosons. This suggests we should build a bridge between a fully interacting system (the original system) and a partially interacting system (the auxiliary system), and have the replica traverse by exchanging updates. From Fig \ref{fig:atuo_vs_inter_slice}, we observe that winding number measurements in a partially interacting system—where interactions occur only in a subset of imaginary time slices—exhibit significantly weaker correlations compared to those in a fully interacting system, where all imaginary time slices are interacting. Therefore, we define the number of interacting slices in the configuration as $p'$. In contrast, we find that energy measurements are less affected because the influence of the global topology of ring polymers is small. Consequently, conventional Monte Carlo updates are sufficient for accurate energy estimation.

\begin{figure}[tbp]
    \centering
    \includegraphics[width=0.5\textwidth]{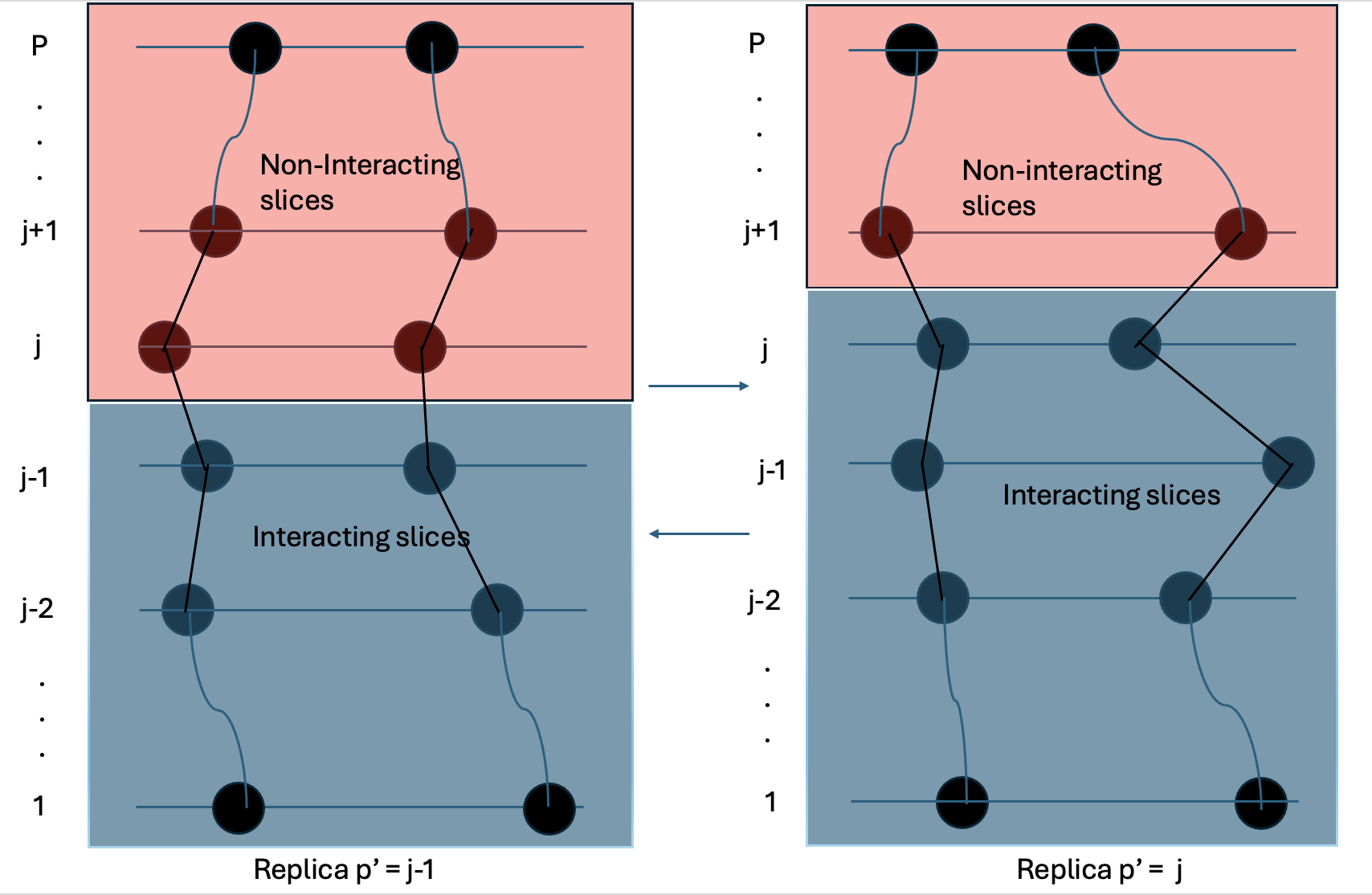}
    \caption{Exchange update between replica index $j$ and $j-1$, indicating the number of interacting slices in the configuration. If the proposal is accepted, replica 1(left replica) will turn on one more interacting slice, and replica 2(right replica) will turn off one more interacting slice. Curved lines represent an abbreviated path to the end of the worldlines.}
    \label{fig:emc_fig}
\end{figure}

In the following, we explain the details of the exchange update between replicas at partially interacting systems. 
From Eq.~\eqref{eq:partition}, we show that the interactions between beads at each imaginary time slice contribute to the total configuration weight, with their magnitude being proportional to the imaginary time step $\tau$. To ensure a reasonable exchange rate, the interaction must be turned off gradually. We therefore choose to turn off slice-by-slice interactions to prepare a sequence of replicas.

We consider the simple Metropolis-Hastings criterion for two replicas, with $p'_1 = j-1$ and $p'_2=j$ (denoted by replica 1 and 2). The acceptance probability of the exchange update of these two replicas will be
\begin{equation}
    P_{j} = \min\{1, \frac{e^{-\tau V(\mathbf{x}_j^{(1)})}}{e^{-\tau V(\mathbf{x}_j^{(2)})}}\}
    \label{emc_exchange}
\end{equation}
where $\mathbf{x}_j^{(1)}$  and $\mathbf{x}_j^{(2)}$ denote configurations at imaginary slice $j$ in replica 1 and 2, respectively. After a successful exchange update, interactions at slice $j$ of replica 1 will be turned on, while those of replica 2 will be turned off as shown in Fig.~\ref{fig:emc_fig}. In the framework of stochastic potential switching, we also need to account for the bond configuration to satisfy the detailed balance condition.One straightforward approach is to keep the acceptance probability in Eq.~\eqref{emc_exchange}, i.e., to perform the exchange test using the marginal positional weight in which the SPS bond variables are integrated out. After an accepted exchange, we update the SPS bond configuration only on the slice whose interaction status is changed: for the replica in which slice $j$ becomes noninteracting, we erase all SPS bonds on that slice; for the replica in which slice $j$ becomes interacting, we create bonds independently for each pair $(a,b)$ on that slice by a heat-bath update. In practice, the bond between $(a,b)$ is created with probability
\begin{equation}
    \exp\!\left[-\tau\bigl(\bar U(r_{ab})-U(r_{ab})\bigr)\right],
\end{equation}
and otherwise no bond is assigned, so that the bond configuration on the newly activated slice is drawn from the exact SPS conditional distribution at fixed positions. This Metropolis step followed by a heat-bath bond refresh satisfies detailed balance in the extended ensemble while retaining the simple acceptance rule in Eq.~\eqref{emc_exchange}. Although the bond refresh scales as $O(N^2)$ when performed naively, it is executed only when a slice is switched on/off (once per time slice along the replica walk), and its overhead is negligible compared with the repeated local updates of bead positions.
In this way, we slightly open the window into the non-interacting region, ensuring more frequent swap updates for ring polymers. From Eq.~\eqref{emc_exchange}, we note that because the energy difference is an extensive quantity, the acceptance rate decreases as the system size increases, an issue addressed in a later section. 

\begin{figure}[tbp]
    \centering
    \includegraphics[width=0.5\textwidth]{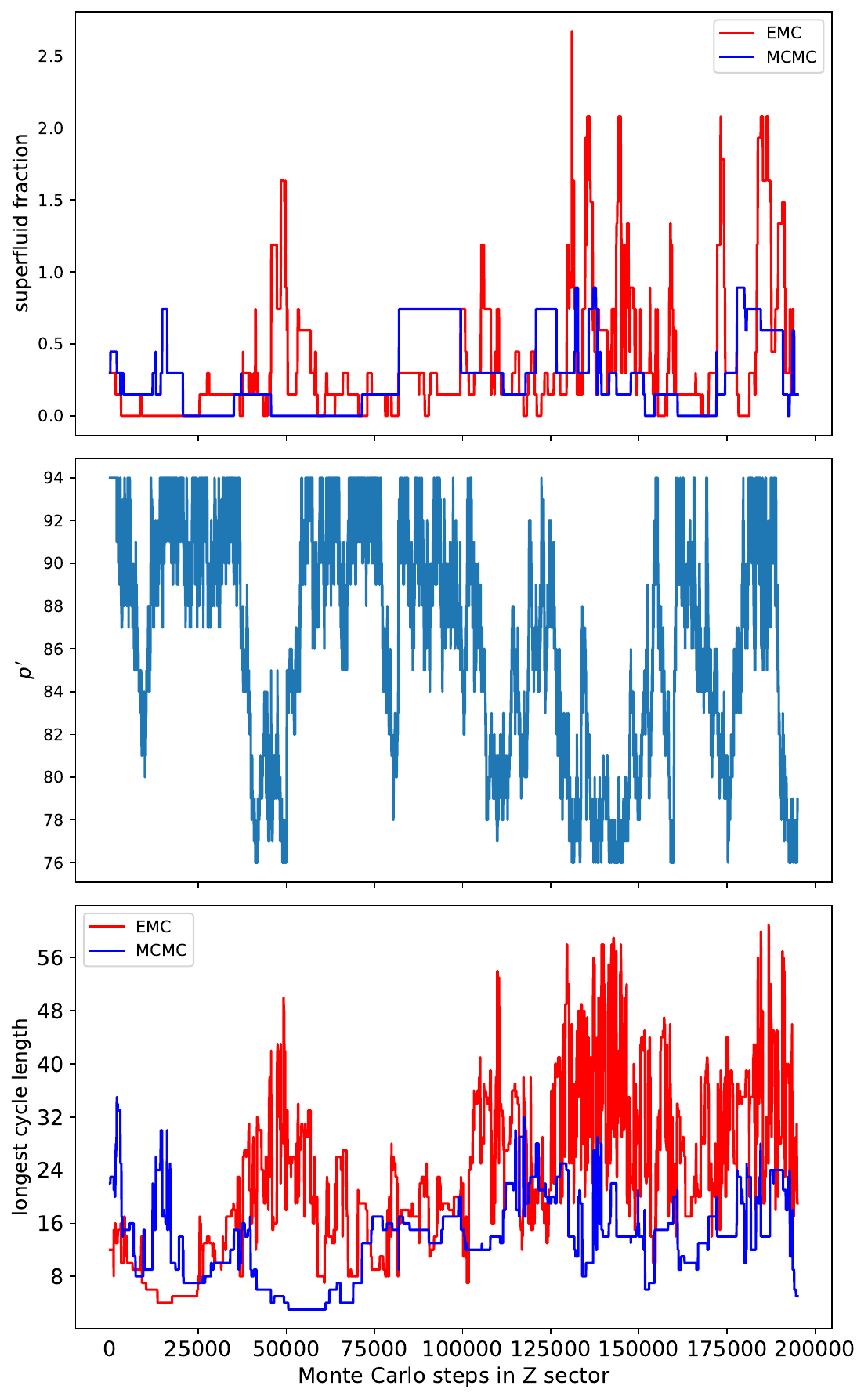}
    \caption{Trajectory of replica at system size $N = 128$ as it traverses between fully interacting system($p' = P = 94$) and partially interacting system ($p' = 76$). Measurements of superfluid fraction and longest cycle length are compared with the MCMC simulation, which has only one replica staying at $p' = 94$. }
    \label{fig:trajectory}
\end{figure}

In Fig \ref{fig:trajectory}, we present simulation results for \( N = 128 \) using 19 replicas for exchange updates. At \( 2.12 \)K, which is slightly above the critical temperature of superfluid $^4$He, the system is simulated with 94 imaginary time slices, using the default imaginary time step \( \tau \).  

In the middle panel of the figure, the blue path represents a typical trajectory of a single replica during the simulation. This replica starts in the fully interacting regime (replica index 94) and walks randomly between indices 94 and 76, which is the lower limit of the number of interacting slices in the simulation.  The horizontal axis corresponds to the Monte Carlo steps of this replica in the Z sector, where all ring polymers are closed during worm updates. The top and bottom plots show measurements of the superfluid fraction and the longest ring-polymer cycle for this replica. These results are compared with those obtained using a single MCMC algorithm without exchange updates.  

We observe that the dynamics of configurations staying in the fully interacting regime (MCMC) are significantly slower than those of replicas undergoing exchange updates. For superfluid fraction, which depends on the worldlines' winding number (see Eq.~\eqref{eq:superfluid_fraction} below), the blue curve in the top panel exhibits long horizontal segments, indicating that the configuration remains trapped in the same winding number sector for an extended number of Monte Carlo steps. In contrast, replicas in Exchange Monte Carlo (EMC) can explore new winding-number sectors more frequently, particularly as they approach lower replica indices, near MC steps such as 50,000 or 135,000.

In the bottom panel, we also observe that the longest ring polymer in the configuration fluctuates more frequently in EMC, particularly in the lower replica-index region. This behavior arises because, within the PIMC framework, beads at neighboring imaginary time slices experience strong spring-like interactions, preventing them from moving far within just a few imaginary time slices. Consequently, for a ring polymer to wind around the periodic box—e.g., from \( x \) to \( x + L \)—a sufficiently long imaginary time is required for it to propagate a distance \( L \). The ability to efficiently sample configurations with large permutation cycles is crucial for accurately measuring the superfluid fraction.

\subsubsection{Nonuniform distribution of imaginary time slices}
In the original Exchange Monte Carlo method, determining the distribution of replicas in temperature space plays a crucial role in improving simulation performance~\cite {rathore2005optimal,hamze2010robust}. A simple approach is to use a geometric distribution of temperatures, ensuring that the swap probability remains constant across all temperatures and is sufficiently high to guarantee a round trip for each replica.  

\begin{figure}[tbp]
    \centering
    \includegraphics[width=0.50\textwidth]{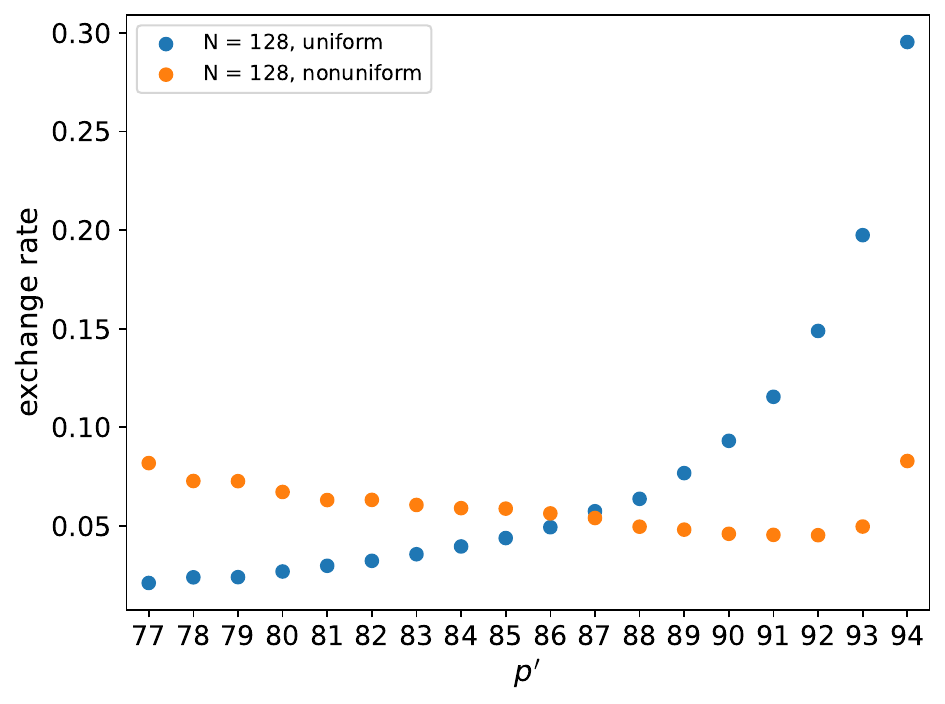}
    \caption{Exchange rate at different $p'$ for uniform and nonuniform distribution of imaginary time slices. After modeling the slice distribution, the exchange rates are flattened, and the smallest exchange rate is increased.}
    \label{fig:exchange_rate_non_uniform}
\end{figure}

In the framework of PIMC, the factor that determines the exchange rate is the imaginary time step \( \tau \) rather than temperature. From Eq.~\eqref{emc_exchange}, one might assume that a uniform distribution of the imaginary time step \( \tau \) would result in an approximately uniform distribution of exchange rates. However, as shown by the blue data points in Fig.~\ref{fig:exchange_rate_non_uniform}, the larger the non-interacting region, the more difficult it is for the exchange update to be accepted. The reason for this is the strong spring-like interactions between beads at neighboring imaginary time slices, which cause the bead distribution at neighboring slices to be correlated. If the interaction is turned off at just one imaginary time slice, the bead distribution at that slice is still constrained by the two adjacent slices, maintaining a high overlap of energy with interacting slices. In contrast, on the left side of the figure, a larger region where interactions are turned off allows ring polymers to overlap, leading to much lower energy overlap with the interacting configuration. 

\begin{figure}[tbp]
    \centering
    \includegraphics[width=0.5\textwidth]{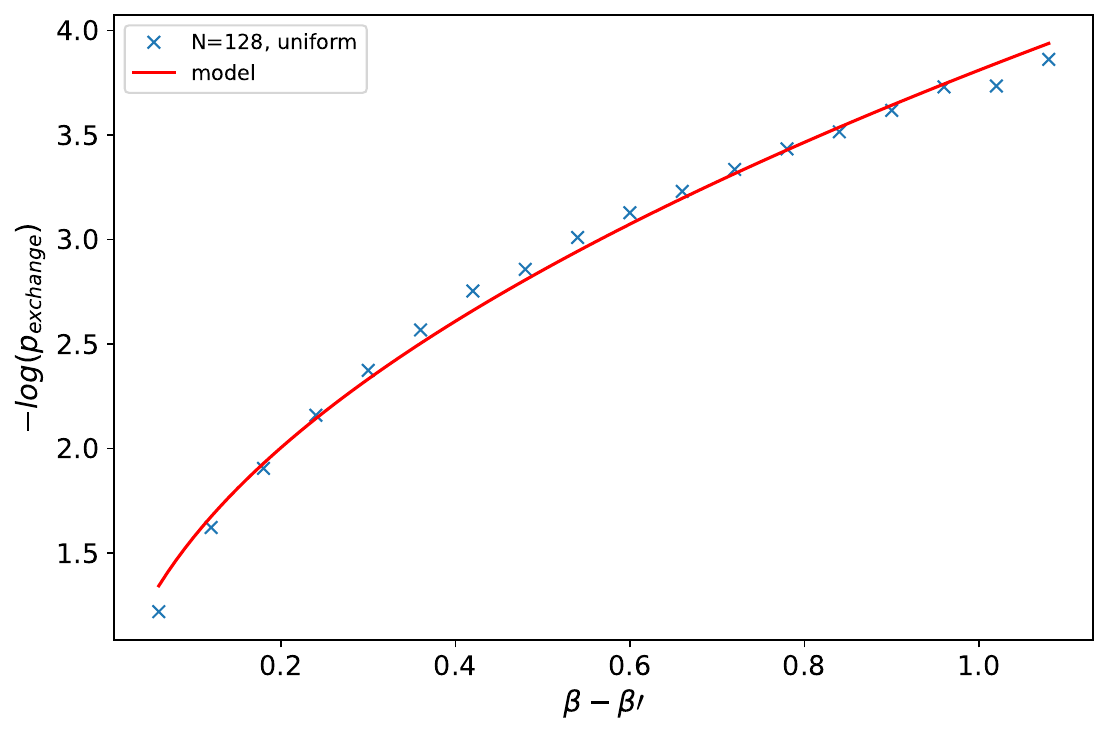}
    \caption{Negative log of exchange rate vs. imaginary time of non-interacting region for system size \(N = 128\). Model ansatz is assumed to be $f(\beta - \beta') = A \sqrt{\beta - \beta'} + B.$}
    \label{fig:modelling}
\end{figure}

To obtain a uniform distribution of the exchange rate, we have to consider non-uniform distribution of imaginary time step. We thus modify our Suzuki-Trotter decomposition: 
\begin{multline}
    \exp{-\tau_j(\hat{T} + \hat{V})} \\ \approx\exp{-\frac{\tau_j\hat{V}}{2}}\exp{-\tau_j\hat{T}}\exp{-\frac{\tau_j\hat{V}}{2}}.
\end{multline}
Accordingly, the weight of interaction at slice $j$ changes from $e^{-\tau V(\mathbf{x}_j)}$ to $e^{-\frac{1}{2}(\tau_j + \tau_{j-1}) V(\mathbf{x}_j)}$.
We propose that the exchange rate between replica that has $j$ interacting slices and $j-1$ interacting slices is
\begin{equation}
p_j \sim \exp\!\left[-\frac{\tau_j + \tau_{j-1}}{2}\, f(\beta - \beta'_j)\right].
\end{equation}

where $\tau_j$ is the imaginary time step of slice \(j\). $\beta - \beta'_j$ represents the imaginary time of non-interacting region,  and $f$ is an ansatz function. Using blue data points from a uniform distribution in Fig \ref{fig:exchange_rate_non_uniform}, we can fit the ansatz function f shown in Fig \ref{fig:modelling}. In this way, we can obtain a set of $\tau$ that flattens the exchange probability by solving the system of equations below:

\begin{subequations}\label{eq:system}
\begin{align}
\tag{\theequation a}\label{eq:system_a}
\sum_{j=p'_0+1}^{P}\frac{1}{2}(\tau_j+\tau_{j-1})
= \beta - \beta'_0, \\[2pt]
\frac{1}{2}(\tau_j+\tau_{j-1})\, f\!\Bigl(\beta - \beta'_0
 - \sum_{p'_0+1 \le k < j}\frac{1}{2}(\tau_k+\tau_{k-1})\Bigr)
&= C, \tag{\theequation b}\label{eq:system_b}
\intertext{\centering\notag for all {j} such that $\,p'_0+1 \le j \le P,$}
\end{align}
\end{subequations}
where \( \beta'_0 \) is the total interacting imaginary time of the replica that has smallest interacting imaginary time, $p'_0$ is the largest interacting slice in that replica, and \( C \) is a constant. As shown by the orange data points in Fig.~\ref{fig:exchange_rate_non_uniform}, we successfully obtain a more flattened distribution of exchange rates while keeping the overall interacting region of imaginary time slices, \( \beta^{\prime}_0 \), unchanged. The rightmost data point shows an anomaly likely due to edge effects when solving the system of equations, since the fully interacting replica has no non-interacting region. 

\begin{figure}[tbp]
    \centering
    \includegraphics[width=0.50\textwidth]{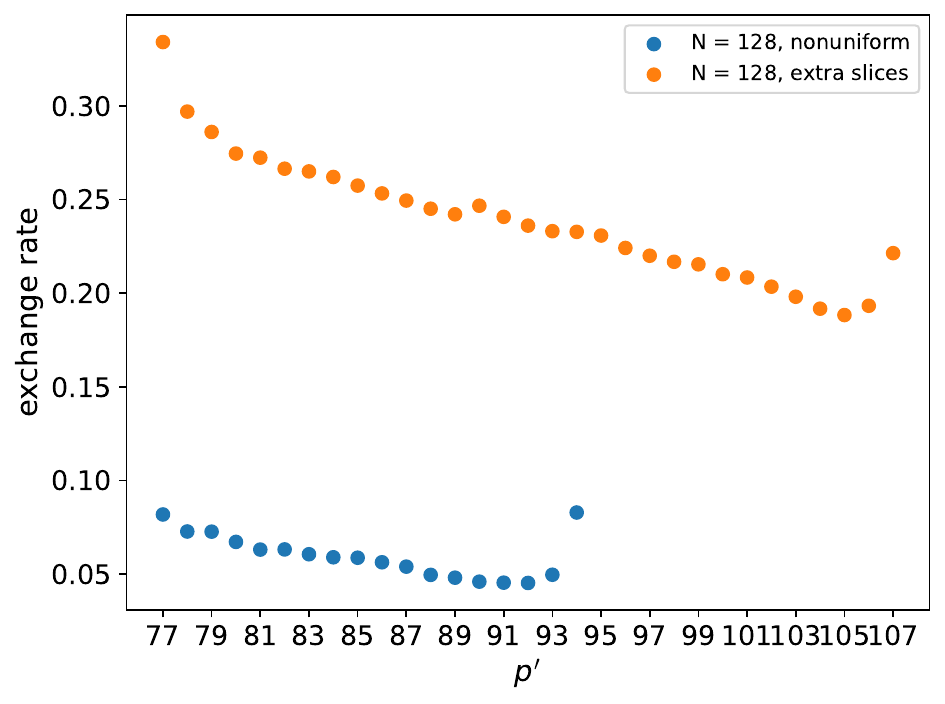}
    \caption{Exchange rate at different $p'$ after inserting extra slices. The interacting region $\beta'_0$ and the number of interacting slices $p'_0$ of the lowest replica index stay the same. The extra slice is inserted only in the non-interacting region. The lowest exchange rate after inserting extra slices becomes nearly $20\%$, ensuring a feasible round-trip of replicas.}
    \label{fig:exchange_rate_extra}
\end{figure}

Another issue is that the overall exchange rates remain low, which can only be improved by adding more imaginary time slices to the non-interacting region. As the spring-like interaction becomes stronger with an increasing number of imaginary time slices, the energy overlap will become larger, leading to an increase in the exchange rate. One might assume that as the exchange rate for each pair of replicas increases, the distance between the two extremes also increases, resulting in the same overall round-trip time, and thus the simulation will not improve. This statement is partially correct: the trade-off exists: the smaller each step exchange update is (though with an increased acceptance probability), the longer it takes for one replica to finish a round-trip. One cannot infinitely increase the efficiency of the exchange update by setting replicas to infinitesimally different values. We here follow the general rule of $20\%$~\cite{katzgraber2006feedback}, and adjust the number of replicas in the simulation as shown in the Fig.~\ref {fig:exchange_rate_extra}. As a result, the overall number of round-trips still increases by a factor of 1.5 due to a larger number of replicas.

\section{PHYSICAL OBSERVABLES}
In the current Exchange Monte Carlo framework, only the largest replica index, which represents the fully interacting system, is of interest. Thus, we only measure when the replica returns to the largest index. 
\subsection{Energy}
For a nonuniform distribution of imaginary time in PIMC, the expectation value of the energy is given by
\begin{equation}
    E = -\frac{1}{Z}\frac{dZ}{d\beta} = -\frac{1}{Z}\sum_{i}\frac{\partial Z}{\partial\tau_i}\frac{d\tau_i}{d\beta},
\end{equation}
where $Z$ is the partition function and $\tau_i$ is the imaginary time step. Here, the nonuniform discretization $\{\tau_i\}_{i=1}^P$ is first fixed at the target inverse temperature $\beta$
(e.g., by solving Eqs.\eqref{eq:system} to flatten the exchange rates) and then treated as part of the discretization scheme.
To define the thermodynamic derivative $E=-\partial_\beta \ln Z$ unambiguously, we take the derivative at fixed
$a_i \equiv \tau_i/\beta$, i.e.,
\begin{equation}
    \tau_i(\beta)=\beta\,a_i, \qquad \sum_{i=1}^P a_i = 1,
\end{equation}
so that
\begin{equation}
    \frac{d\tau_i}{d\beta}=a_i=\frac{\tau_i}{\beta}.
\end{equation}
Any residual dependence on the particular choice of $\{a_i\}$ is a finite $P$ discretization error and
vanishes in the limit $P\to\infty$.
. The estimator can be further derived as
\begin{multline}
    E_\text{mc} = \frac{PDN}{2\beta} - \sum_{i=a}^{N}\sum_{j=1}^{P} \frac{\|\mathbf{x}_{j+1,a} - \mathbf{x}_{j,a}\|^2}{2\tau_j\beta} \\
    + \sum_{j=1}^{P} \frac{\tau_j + \tau_{j-1}}{2\beta} E_\text{inter}(\mathbf{x}_{j}),
\end{multline}
where \(E_\text{inter}(\mathbf{x}_{j})= \sum_{1\le a<b\le N} U\!\left(|\mathbf{x}_{j,a}-\mathbf{x}_{j,b}|\right)
\quad \).

In the context of the stochastic potential switching, the first two terms of the estimator stay the same. The third term, which accounts for potential interactions, can be measured by using the following estimator:
\begin{multline}
    E_\text{inter}(\mathbf{x}_j) = \sum_{a < b}\Big[ (1-d_{a,b})\tilde{U}(\mathbf{x}_{j,a},\mathbf{x}_{j,b}) \\ + d_{a,b}
    \frac{U(\mathbf{x}_{j,a},\mathbf{x}_{j,b})-S\,\tilde{U}(\mathbf{x}_{j,a},\mathbf{x}_{j,b})}{1-S}\Big],
\end{multline}
where $S$ is the switching probability as a function of the distance between particles, and $d_{a,b}$ is 1 if two particles share interaction $\bar{U}$, where $\bar{U}$ is the potential energy if the switching fails as explained in Section II. This estimator can be performed in $O(NP)$ using the binning technique for the long-range potential. 

\subsection{Superfluid fraction}
In a periodic simulation cell, the superfluid response is encoded in the topological winding of
bosonic worldlines around the boundaries. Let \(\mathbf{W}=(W_1,\ldots,W_D)\) denote the
\(D\)-component winding-number vector, where each component \(W_\mu\in\mathbb{Z}\) counts the net number
of times the worldlines wrap around the simulation box in direction \(\mu\). We also define
\begin{equation}
    W^2 \equiv \sum_{\mu=1}^{D} W_\mu^2 .
\end{equation}
The estimator of the superfluid fraction can be given by ~\cite{pollock1987path,krauth2006statistical}
\begin{equation}
    \frac{\rho_s}{\rho}
    = \frac{L^2}{D\,N\,\beta}\,\langle W^2\rangle ,
    \label{eq:superfluid_fraction}
\end{equation}
where \(L\) is the linear system size (assumed isotropic), \(N\) is the particle number, $\rho$  is the density given by $N / L^D$ and \(\beta\) is
the inverse temperature.

\subsection{Condensate fraction}
Condensate fraction can be estimated by the ratio between the number of G sector (where there is one cut along the path) and Z sector (original configuration without cut)~\cite{boninsegni2006worm}.
\begin{equation}
    n_0 = \frac{1}{C_1' PVN }\frac{\langle \delta_G \rangle}{\langle \delta_Z \rangle},
\end{equation}
where the notations are the same as in Eq.~\eqref{open update}. In the case of a nonuniform distribution of imaginary time slices, we should fix the imaginary time rather than the number of slices when considering open and closed updates for transitions between the Z sector and the G sector. In other words, we replace \(m\tau\) by the accumulated imaginary-time gap \(\Delta\tau = \sum_{k=0}^{m-1} \tau_{j+k}\) in the free-particle propagator factors. Moreover, it is better to automatically tune $C_1^{\prime}$ to achieve a balanced number of Z and G sectors at each replica index, thereby improving mixing. In the present work, we adopt Wang-Landau's sampling~\cite{landau2004new} during the thermalization phase to find the proper $C_1^{\prime}$. 

\section{NUMERICAL RESULTS}
In a Markov Chain Monte Carlo simulation, one of the most important criteria for assessing effectiveness is the autocorrelation time.  We apply binning analysis~\cite{ambegaokar2010estimating} to accurately estimate the statistical uncertainty of the measurements. 

In binning analysis, we divide the time series into several bins of size \( k \). If there are \( M \) bins in total, the standard error of the mean for these bins is given by  
\begin{equation}
    \sigma_k = \sqrt{\frac{1}{M} \frac{1}{M-1} \sum_{i=1}^{M} (O_i^{(k)} - \bar{O})^2},
\end{equation}
where \( O_i^{(k)} \) represents the mean of the observable in bin index \( i \), and \( \bar{O} \) denotes the mean of the entire time series. We gradually increase the bin size \( k \) until the variance or standard error of the mean stabilizes. The autocorrelation time of the observable can then be estimated as  
\begin{equation}
    \tau_{int} = \frac{1}{2} \left( \frac{\sigma_{\infty}^2}{\sigma_1^2} - 1 \right)
\end{equation}
where \( \sigma_{\infty}^2 \) is the variance of the bins when it becomes stabilized.
\begin{figure}[tbp]
    \centering
    \includegraphics[width=0.5\textwidth]{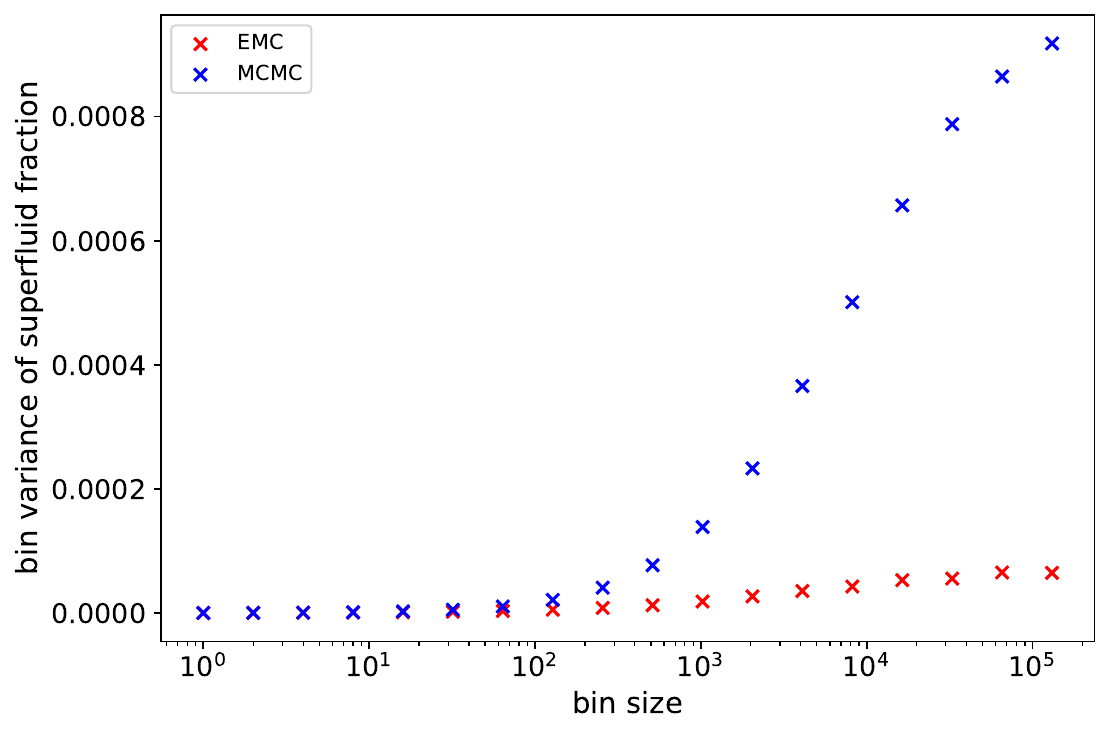}
    \caption{Binning analysis of superfluid fraction for EMC and MCMC at system size N=32 and near critical temperature.}
    \label{fig:binning}
\end{figure}
As shown in Fig.~\ref{fig:binning}, we plot the variance as a function of bin size ranging from 1 to \( 2^{17} \) for the superfluid fraction until it stabilizes. We observe that the variance of the mean plateaus at a bin size of around \(10^5\). The standard error obtained using MCMC is nearly four times larger than that obtained using EMC, demonstrating EMC's superior performance in reducing autocorrelation times, which can be attributed to frequent exchange updates and rapid exploration of the configuration at small replica indices. 

We also comment on the computational efficiency of the present EMC scheme. 
Since EMC uses multiple replicas, it is useful to distinguish the total serial cost from the parallel wall-clock cost. 
Let \(R\) be the number of replicas and let \(c_r\) denote the cost of one update cycle of replica \(r\). 
If all replicas are propagated serially, the cost of one EMC cycle is approximately
\[
C_{\rm serial}^{\rm EMC} \simeq \sum_{r=1}^{R} c_r + C_{\rm ex},
\]
where \(C_{\rm ex}\) is the cost of the exchange attempts. 
In the parallel implementation used here, however, the replicas are propagated independently between exchange attempts and are assigned to different cores. 
The corresponding wall-clock cost is therefore
\[
C_{\rm wall}^{\rm EMC} \simeq \max_{r} c_r + C_{\rm ex} + C_{\rm comm},
\]
where \(C_{\rm comm}\) denotes the communication and synchronization overhead. 
For the present choice of replicas, the fully interacting target replica has the largest computational cost, while auxiliary replicas with fewer interacting imaginary-time slices are cheaper. 
Thus \(\max_r c_r\) is approximately the cost of an ordinary single-replica PIMC update of the target system, and the wall-clock cost of one EMC cycle is comparable to that of conventional MCMC, up to the relatively small exchange and communication overheads. 
Consequently, the observed reduction of the integrated autocorrelation time for permutation- and winding-number-sensitive observables translates into a reduction of the parallel wall-clock time required to obtain effectively independent samples. 
This parallel use of auxiliary replicas is also different from running \(R\) independent MCMC chains. 
Independent chains can reduce statistical error by averaging, but they do not reduce the intrinsic tunneling time among permutation and winding-number sectors in each chain. 
In contrast, EMC modifies the sampling dynamics of the target replica by allowing configurations to visit auxiliary replicas in which world-line rearrangements are less strongly constrained by interparticle interactions.

\begin{figure}[tbp]
    \centering
    \includegraphics[width=0.5\textwidth]{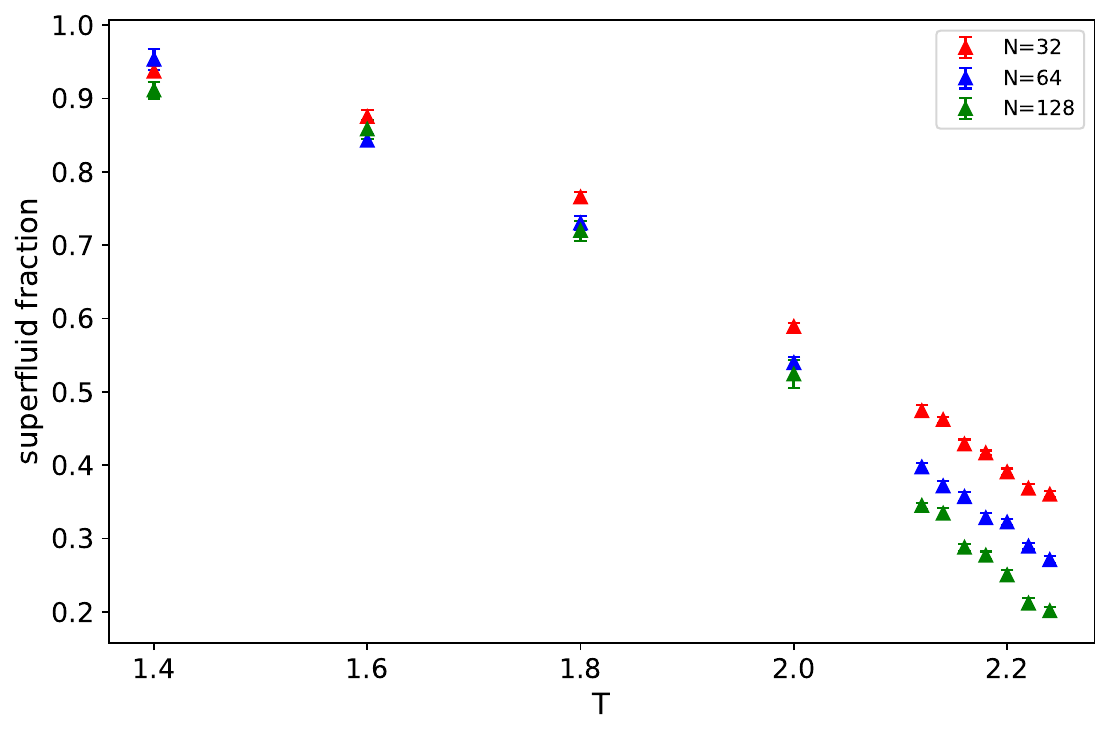}
    \caption{Superfluid fraction at different system sizes. Error bars are obtained from ten independent Monte Carlo runs.}
    \label{fig:obs_super}
\end{figure}

\begin{figure}[tbp]
    \centering
    \includegraphics[width=0.5\textwidth]{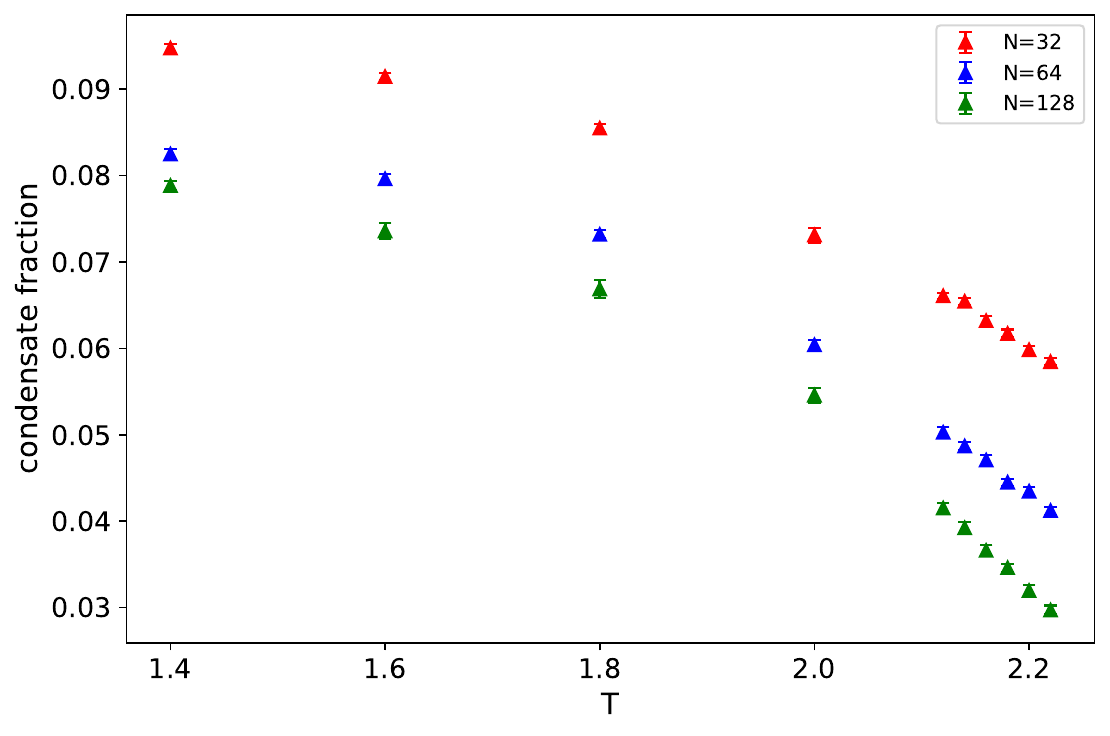}
    \caption{Condensate fraction at different system sizes. Error bars are obtained from ten independent Monte Carlo runs.}
    \label{fig:obs_conden}
\end{figure}

\begin{figure}[tbp]
    \centering
    \includegraphics[width=0.5\textwidth]{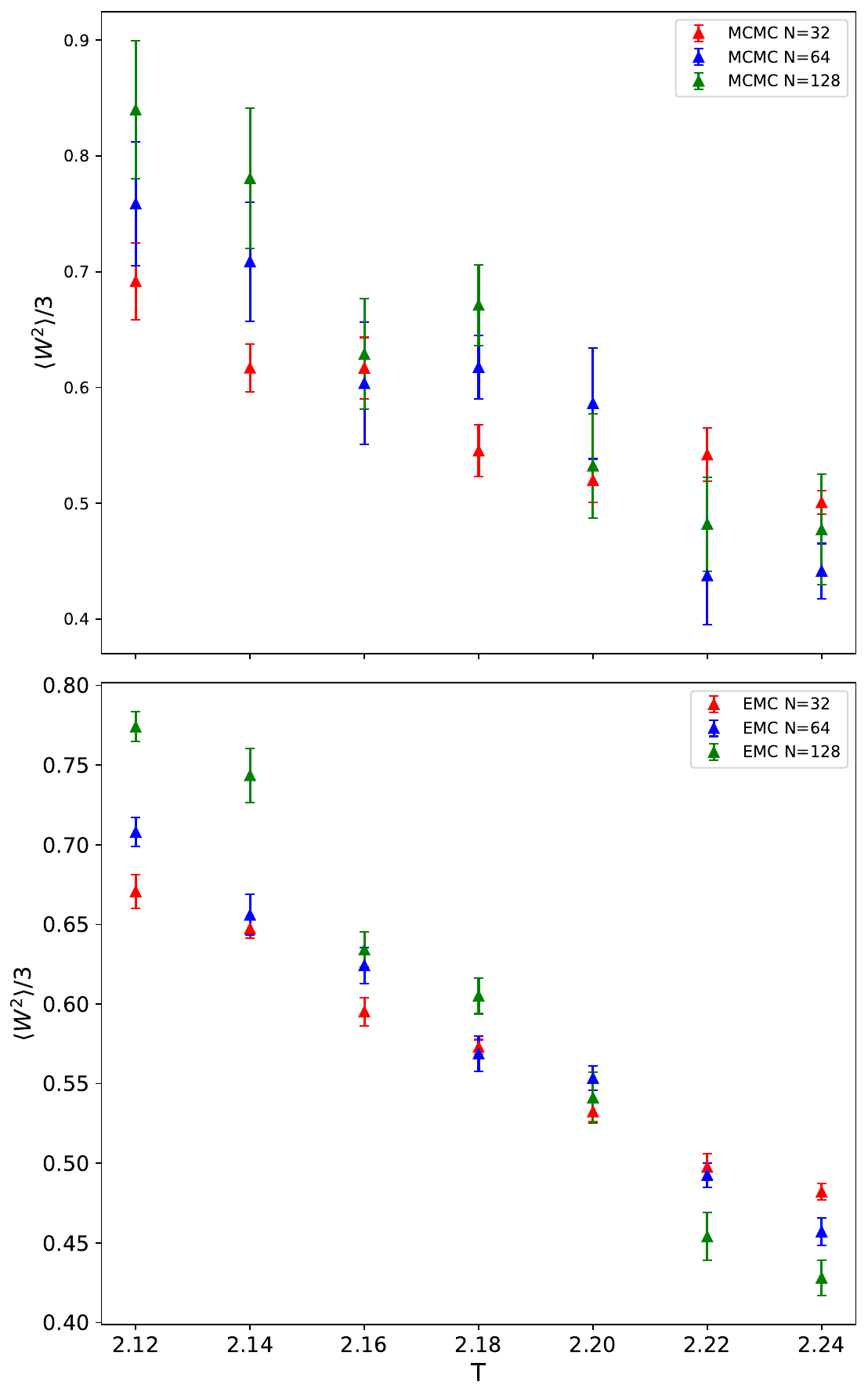}
    \caption{Winding number squared at different system sizes from MCMC simulation and EMC simulation. Error bars are obtained from ten independent Monte Carlo runs. The crossing point is scale invariant at the critical temperature.}
    \label{fig:obs_winding}
\end{figure}

In Figs \ref{fig:obs_super} and \ref{fig:obs_conden}, we present results for the superfluid fraction and condensate fraction as a function of temperature for different system sizes (at density $\rho = 0.02198\,\mathrm{\AA}^{-3} $ and $D=3$). To extract the critical exponent and obtain the critical temperature, we perform finite-size scaling analysis using the scaling form from~\cite{boninsegni2006worm}, 

\begin{equation}
    \frac{\langle W^2 \rangle}{3} = f(|T-T_c|L^{1/\nu}),
\end{equation}
which implies that the winding number should be invariant across all system sizes at the critical temperature. In Fig.~\ref{fig:obs_winding}, compared to the MCMC simulation, the reduced error bar measured by the EMC method can help us determine the crossing point much more accurately. Applying Bayesian Scaling Analysis~\cite{Harada2011}, we obtain the critical temperature around 2.195K, which is highly consistent with the result of the worm algorithm~\cite{boninsegni2006worm}.

\section{DISCUSSION AND CONCLUSION}
In this paper, we have introduced a novel Exchange Monte Carlo (EMC) method applied to the continuous-space Path Integral Monte Carlo (PIMC) model. We have redefined the exchange update mechanism to operate along the axis of the number of interacting slices, rather than temperature, which was the basis of the original EMC. Our approach demonstrates that introducing auxiliary replicas across different interacting regimes can significantly accelerate Monte Carlo dynamics, improving the original replica's sampling efficiency. Furthermore, we explore the relationship between the exchange rate and the distribution of imaginary time steps and show that using a nonuniform time-step distribution and inserting extra slices in the non-interacting region can flatten and enhance the exchange rate.

In addition, we introduce the Stochastic Potential Switching (SPS) framework for treating interatomic interactions, which reduces the computational complexity from $O(N^2)$ to $O(N)$. The smooth decomposition of interaction weights in the SPS method ensures compatibility with any Hamiltonian Monte Carlo (HMC) update, thereby improving efficiency and generalizability in simulations.

The mechanism underlying this improvement is more general than the particular
binary choice of interacting and noninteracting slices used in this work.
In continuous-space bosonic PIMC, slow winding-number sampling reflects a
competition between two effects. The kinetic, or spring-like, terms along the
imaginary-time direction require a sufficiently long imaginary-time interval for
a world line to diffuse across the periodic simulation cell and form large
permutation cycles. At the same time, the equal-time interparticle interactions
within each imaginary-time slice suppress large rearrangements of beads, in
particular for realistic pair potentials with strong short-range repulsion.
The present exchange coordinate weakens one side of this competition by
introducing auxiliary replicas in which the pair interaction is removed from
part of the imaginary-time interval. These replicas allow world lines to
rearrange more easily, while exchange moves transfer the resulting improved
sampling of permutation cycles and winding-number sectors back to the fully
interacting target replica.

This viewpoint suggests several concrete future applications and extensions.
The method should be particularly useful for continuous-space bosonic PIMC
simulations with realistic interatomic pair potentials, where strong pair
repulsion suppresses global world-line rearrangements. Natural examples include
larger and lower-temperature simulations of superfluid \(^4\)He, other
interacting Bose fluids, Bose mixtures, clusters, films, and confined or
inhomogeneous geometries~\cite{Sindzingre1989HeClusters,Gordillo1998HeliumFilm}. In these systems, the central difficulty is not only
the cost of evaluating the potential, but also the conflict between the long
imaginary-time propagation needed for winding-number changes and the high
rejection probability caused by equal-time pair interactions. A promising
algorithmic extension is to replace the present on/off activation of pair
interactions by a continuous interaction-strength annealing scheme. For example,
one may introduce auxiliary replicas with actions of the form
\[
S_r[\mathbf R]
=
S_{\rm kin}[\mathbf R]
+
\sum_{l=1}^{M} \Delta\tau_l \lambda_{r,l}
V(\mathbf R_l),
\qquad
0 \leq \lambda_{r,l} \leq 1,
\]
where \(V(\mathbf R_l)\) denotes the pair-interaction energy at imaginary-time
slice \(l\), and the physical target replica corresponds to
\(\lambda_{r,l}=1\) for all \(l\). The present method can be regarded as the
binary case \(\lambda_{r,l}\in\{0,1\}\), while a continuous schedule would
interpolate more smoothly from weakly interacting auxiliary systems to the
fully interacting system. Such an annealing coordinate may improve the overlap
between neighboring replicas and can be optimized using exchange rates,
round-trip times, or the wall-clock time required to obtain independent samples
of winding-number-sensitive observables.

\begin{acknowledgments}
This work was supported by JSPS KAKENHI Grant Numbers 20H01824 and 24K00543, the Center of Innovation for Sustainable Quantum AI (SQAI), JST Grant Number JPMJPF2221, and JST CREST Grant Number JPMJCR24I1.
XZ thanks the Global Science Graduate Course (GSGC) program for its financial support.
\end{acknowledgments}

%\appendix

%\section{Appendixes}


\begin{thebibliography}{38}%
\makeatletter
\providecommand \@ifxundefined [1]{%
 \@ifx{#1\undefined}
}%
\providecommand \@ifnum [1]{%
 \ifnum #1\expandafter \@firstoftwo
 \else \expandafter \@secondoftwo
 \fi
}%
\providecommand \@ifx [1]{%
 \ifx #1\expandafter \@firstoftwo
 \else \expandafter \@secondoftwo
 \fi
}%
\providecommand \natexlab [1]{#1}%
\providecommand \enquote  [1]{``#1''}%
\providecommand \bibnamefont  [1]{#1}%
\providecommand \bibfnamefont [1]{#1}%
\providecommand \citenamefont [1]{#1}%
\providecommand \href@noop [0]{\@secondoftwo}%
\providecommand \href [0]{\begingroup \@sanitize@url \@href}%
\providecommand \@href[1]{\@@startlink{#1}\@@href}%
\providecommand \@@href[1]{\endgroup#1\@@endlink}%
\providecommand \@sanitize@url [0]{\catcode `\\12\catcode `\$12\catcode `\&12\catcode `\#12\catcode `\^12\catcode `\_12\catcode `\%12\relax}%
\providecommand \@@startlink[1]{}%
\providecommand \@@endlink[0]{}%
\providecommand \url  [0]{\begingroup\@sanitize@url \@url }%
\providecommand \@url [1]{\endgroup\@href {#1}{\urlprefix }}%
\providecommand \urlprefix  [0]{URL }%
\providecommand \Eprint [0]{\href }%
\providecommand \doibase [0]{https://doi.org/}%
\providecommand \selectlanguage [0]{\@gobble}%
\providecommand \bibinfo  [0]{\@secondoftwo}%
\providecommand \bibfield  [0]{\@secondoftwo}%
\providecommand \translation [1]{[#1]}%
\providecommand \BibitemOpen [0]{}%
\providecommand \bibitemStop [0]{}%
\providecommand \bibitemNoStop [0]{.\EOS\space}%
\providecommand \EOS [0]{\spacefactor3000\relax}%
\providecommand \BibitemShut  [1]{\csname bibitem#1\endcsname}%
\let\auto@bib@innerbib\@empty
%</preamble>
\bibitem [{\citenamefont {Landau}\ and\ \citenamefont {Binder}(2021)}]{landau2021guide}%
  \BibitemOpen
  \bibfield  {author} {\bibinfo {author} {\bibfnamefont {D.~P.}\ \bibnamefont {Landau}}\ and\ \bibinfo {author} {\bibfnamefont {K.}~\bibnamefont {Binder}},\ }\href@noop {} {\emph {\bibinfo {title} {A Guide to {Monte Carlo} Simulations in Statistical Physics}}}\ (\bibinfo  {publisher} {Cambridge University Press},\ \bibinfo {year} {2021})\BibitemShut {NoStop}%
\bibitem [{\citenamefont {Liu}(2001)}]{liu2001monte}%
  \BibitemOpen
  \bibfield  {author} {\bibinfo {author} {\bibfnamefont {J.~S.}\ \bibnamefont {Liu}},\ }\href@noop {} {\emph {\bibinfo {title} {{Monte Carlo} Strategies in Scientific Computing}}}\ (\bibinfo  {publisher} {Springer},\ \bibinfo {year} {2001})\BibitemShut {NoStop}%
\bibitem [{\citenamefont {Brooks}(1998)}]{brooks1998markov}%
  \BibitemOpen
  \bibfield  {author} {\bibinfo {author} {\bibfnamefont {S.}~\bibnamefont {Brooks}},\ }\bibfield  {title} {\bibinfo {title} {Markov chain {Monte Carlo} method and its application},\ }\href@noop {} {\bibfield  {journal} {\bibinfo  {journal} {J. R. Stat. Soc. Ser. D (The Statistician)}\ }\textbf {\bibinfo {volume} {47}},\ \bibinfo {pages} {69} (\bibinfo {year} {1998})}\BibitemShut {NoStop}%
\bibitem [{\citenamefont {Eraker}(2001)}]{eraker2001mcmc}%
  \BibitemOpen
  \bibfield  {author} {\bibinfo {author} {\bibfnamefont {B.}~\bibnamefont {Eraker}},\ }\bibfield  {title} {\bibinfo {title} {{MCMC} analysis of diffusion models with application to finance},\ }\href@noop {} {\bibfield  {journal} {\bibinfo  {journal} {J. Bus. Econ. Stat.}\ }\textbf {\bibinfo {volume} {19}},\ \bibinfo {pages} {177} (\bibinfo {year} {2001})}\BibitemShut {NoStop}%
\bibitem [{\citenamefont {Pollock}\ and\ \citenamefont {Ceperley}(1987)}]{pollock1987path}%
  \BibitemOpen
  \bibfield  {author} {\bibinfo {author} {\bibfnamefont {E.~L.}\ \bibnamefont {Pollock}}\ and\ \bibinfo {author} {\bibfnamefont {D.~M.}\ \bibnamefont {Ceperley}},\ }\bibfield  {title} {\bibinfo {title} {Path-integral computation of superfluid densities},\ }\href@noop {} {\bibfield  {journal} {\bibinfo  {journal} {Phys. Rev. B}\ }\textbf {\bibinfo {volume} {36}},\ \bibinfo {pages} {8343} (\bibinfo {year} {1987})}\BibitemShut {NoStop}%
\bibitem [{\citenamefont {Sindzingre}\ \emph {et~al.}(1989{\natexlab{a}})\citenamefont {Sindzingre}, \citenamefont {Klein},\ and\ \citenamefont {Ceperley}}]{sindzingre1989path}%
  \BibitemOpen
  \bibfield  {author} {\bibinfo {author} {\bibfnamefont {P.}~\bibnamefont {Sindzingre}}, \bibinfo {author} {\bibfnamefont {M.~L.}\ \bibnamefont {Klein}},\ and\ \bibinfo {author} {\bibfnamefont {D.~M.}\ \bibnamefont {Ceperley}},\ }\bibfield  {title} {\bibinfo {title} {Path-integral {Monte Carlo} study of low-temperature \(^4\)he clusters},\ }\href@noop {} {\bibfield  {journal} {\bibinfo  {journal} {Phys. Rev. Lett.}\ }\textbf {\bibinfo {volume} {63}},\ \bibinfo {pages} {1601} (\bibinfo {year} {1989}{\natexlab{a}})}\BibitemShut {NoStop}%
\bibitem [{\citenamefont {Boninsegni}\ \emph {et~al.}(2006)\citenamefont {Boninsegni}, \citenamefont {Prokof'ev},\ and\ \citenamefont {Svistunov}}]{boninsegni2006worm}%
  \BibitemOpen
  \bibfield  {author} {\bibinfo {author} {\bibfnamefont {M.}~\bibnamefont {Boninsegni}}, \bibinfo {author} {\bibfnamefont {N.~V.}\ \bibnamefont {Prokof'ev}},\ and\ \bibinfo {author} {\bibfnamefont {B.~V.}\ \bibnamefont {Svistunov}},\ }\bibfield  {title} {\bibinfo {title} {Worm algorithm and diagrammatic {Monte Carlo}: A new approach to continuous-space path integral {Monte Carlo} simulations},\ }\href@noop {} {\bibfield  {journal} {\bibinfo  {journal} {Phys. Rev. E}\ }\textbf {\bibinfo {volume} {74}},\ \bibinfo {pages} {036701} (\bibinfo {year} {2006})}\BibitemShut {NoStop}%
\bibitem [{\citenamefont {Hukushima}\ and\ \citenamefont {Nemoto}(1996)}]{hukushima1996exchange}%
  \BibitemOpen
  \bibfield  {author} {\bibinfo {author} {\bibfnamefont {K.}~\bibnamefont {Hukushima}}\ and\ \bibinfo {author} {\bibfnamefont {K.}~\bibnamefont {Nemoto}},\ }\bibfield  {title} {\bibinfo {title} {Exchange {Monte Carlo} method and application to spin glass simulations},\ }\href@noop {} {\bibfield  {journal} {\bibinfo  {journal} {J. Phys. Soc. Jpn.}\ }\textbf {\bibinfo {volume} {65}},\ \bibinfo {pages} {1604} (\bibinfo {year} {1996})}\BibitemShut {NoStop}%
\bibitem [{\citenamefont {Hoffman}\ and\ \citenamefont {Gelman}(2014)}]{hoffman2014no}%
  \BibitemOpen
  \bibfield  {author} {\bibinfo {author} {\bibfnamefont {M.~D.}\ \bibnamefont {Hoffman}}\ and\ \bibinfo {author} {\bibfnamefont {A.}~\bibnamefont {Gelman}},\ }\bibfield  {title} {\bibinfo {title} {The {No-U-Turn} sampler: adaptively setting path lengths in {Hamiltonian Monte Carlo}},\ }\href@noop {} {\bibfield  {journal} {\bibinfo  {journal} {J. Mach. Learn. Res.}\ }\textbf {\bibinfo {volume} {15}},\ \bibinfo {pages} {1593} (\bibinfo {year} {2014})}\BibitemShut {NoStop}%
\bibitem [{\citenamefont {Mak}(2005)}]{mak2005stochastic}%
  \BibitemOpen
  \bibfield  {author} {\bibinfo {author} {\bibfnamefont {C.~H.}\ \bibnamefont {Mak}},\ }\bibfield  {title} {\bibinfo {title} {Stochastic potential switching algorithm for {Monte Carlo} simulations of complex systems},\ }\href@noop {} {\bibfield  {journal} {\bibinfo  {journal} {J. Chem. Phys.}\ }\textbf {\bibinfo {volume} {122}},\ \bibinfo {pages} {214110} (\bibinfo {year} {2005})}\BibitemShut {NoStop}%
\bibitem [{\citenamefont {Suzuki}(1976)}]{suzuki1976generalized}%
  \BibitemOpen
  \bibfield  {author} {\bibinfo {author} {\bibfnamefont {M.}~\bibnamefont {Suzuki}},\ }\bibfield  {title} {\bibinfo {title} {Generalized {Trotter}'s formula and systematic approximants of exponential operators and inner derivations with applications to many-body problems},\ }\href@noop {} {\bibfield  {journal} {\bibinfo  {journal} {Commun. Math. Phys.}\ }\textbf {\bibinfo {volume} {51}},\ \bibinfo {pages} {183} (\bibinfo {year} {1976})}\BibitemShut {NoStop}%
\bibitem [{\citenamefont {Trotter}(1959)}]{trotter1959product}%
  \BibitemOpen
  \bibfield  {author} {\bibinfo {author} {\bibfnamefont {H.~F.}\ \bibnamefont {Trotter}},\ }\bibfield  {title} {\bibinfo {title} {On the product of semi-groups of operators},\ }\href@noop {} {\bibfield  {journal} {\bibinfo  {journal} {Proc. Am. Math. Soc.}\ }\textbf {\bibinfo {volume} {10}},\ \bibinfo {pages} {545} (\bibinfo {year} {1959})}\BibitemShut {NoStop}%
\bibitem [{\citenamefont {Spada}\ \emph {et~al.}(2022)\citenamefont {Spada}, \citenamefont {Giorgini},\ and\ \citenamefont {Pilati}}]{spada2022path}%
  \BibitemOpen
  \bibfield  {author} {\bibinfo {author} {\bibfnamefont {G.}~\bibnamefont {Spada}}, \bibinfo {author} {\bibfnamefont {S.}~\bibnamefont {Giorgini}},\ and\ \bibinfo {author} {\bibfnamefont {S.}~\bibnamefont {Pilati}},\ }\bibfield  {title} {\bibinfo {title} {Path-integral {Monte Carlo} worm algorithm for {B}ose systems with periodic boundary conditions},\ }\href@noop {} {\bibfield  {journal} {\bibinfo  {journal} {Condens. Matter}\ }\textbf {\bibinfo {volume} {7}},\ \bibinfo {pages} {30} (\bibinfo {year} {2022})}\BibitemShut {NoStop}%
\bibitem [{\citenamefont {Bertoin}(1996)}]{bertoin1996levy}%
  \BibitemOpen
  \bibfield  {author} {\bibinfo {author} {\bibfnamefont {J.}~\bibnamefont {Bertoin}},\ }\href@noop {} {\emph {\bibinfo {title} {L{\'e}vy Processes}}}\ (\bibinfo  {publisher} {Cambridge University Press},\ \bibinfo {year} {1996})\BibitemShut {NoStop}%
\bibitem [{\citenamefont {Metropolis}\ \emph {et~al.}(1953)\citenamefont {Metropolis}, \citenamefont {Rosenbluth}, \citenamefont {Rosenbluth}, \citenamefont {Teller},\ and\ \citenamefont {Teller}}]{metropolis1953}%
  \BibitemOpen
  \bibfield  {author} {\bibinfo {author} {\bibfnamefont {N.}~\bibnamefont {Metropolis}}, \bibinfo {author} {\bibfnamefont {A.~W.}\ \bibnamefont {Rosenbluth}}, \bibinfo {author} {\bibfnamefont {M.~N.}\ \bibnamefont {Rosenbluth}}, \bibinfo {author} {\bibfnamefont {A.~H.}\ \bibnamefont {Teller}},\ and\ \bibinfo {author} {\bibfnamefont {E.}~\bibnamefont {Teller}},\ }\bibfield  {title} {\bibinfo {title} {Equation of state calculations by fast computing machines},\ }\href@noop {} {\bibfield  {journal} {\bibinfo  {journal} {J. Chem. Phys.}\ }\textbf {\bibinfo {volume} {21}},\ \bibinfo {pages} {1087} (\bibinfo {year} {1953})}\BibitemShut {NoStop}%
\bibitem [{\citenamefont {Hastings}(1970)}]{hastings1970monte}%
  \BibitemOpen
  \bibfield  {author} {\bibinfo {author} {\bibfnamefont {W.~K.}\ \bibnamefont {Hastings}},\ }\bibfield  {title} {\bibinfo {title} {{Monte Carlo} sampling methods using {Markov} chains and their applications},\ }\href@noop {} {\bibfield  {journal} {\bibinfo  {journal} {Biometrika}\ }\textbf {\bibinfo {volume} {57}},\ \bibinfo {pages} {97} (\bibinfo {year} {1970})}\BibitemShut {NoStop}%
\bibitem [{\citenamefont {Duane}\ \emph {et~al.}(1987)\citenamefont {Duane}, \citenamefont {Kennedy}, \citenamefont {Pendleton},\ and\ \citenamefont {Roweth}}]{duane1987hybrid}%
  \BibitemOpen
  \bibfield  {author} {\bibinfo {author} {\bibfnamefont {S.}~\bibnamefont {Duane}}, \bibinfo {author} {\bibfnamefont {A.~D.}\ \bibnamefont {Kennedy}}, \bibinfo {author} {\bibfnamefont {B.~J.}\ \bibnamefont {Pendleton}},\ and\ \bibinfo {author} {\bibfnamefont {D.}~\bibnamefont {Roweth}},\ }\bibfield  {title} {\bibinfo {title} {Hybrid {Monte Carlo}},\ }\href@noop {} {\bibfield  {journal} {\bibinfo  {journal} {Phys. Lett. B}\ }\textbf {\bibinfo {volume} {195}},\ \bibinfo {pages} {216} (\bibinfo {year} {1987})}\BibitemShut {NoStop}%
\bibitem [{\citenamefont {Betancourt}(2017)}]{betancourt2017conceptual}%
  \BibitemOpen
  \bibfield  {author} {\bibinfo {author} {\bibfnamefont {M.}~\bibnamefont {Betancourt}},\ }\bibfield  {title} {\bibinfo {title} {A conceptual introduction to {Hamiltonian Monte Carlo}},\ }\href@noop {} {\bibfield  {journal} {\bibinfo  {journal} {arXiv}\ } (\bibinfo {year} {2017})},\ \Eprint {https://arxiv.org/abs/1701.02434} {arXiv:1701.02434 [stat.ME]} \BibitemShut {NoStop}%
\bibitem [{\citenamefont {Neal}(2011)}]{neal2011mcmc}%
  \BibitemOpen
  \bibfield  {author} {\bibinfo {author} {\bibfnamefont {R.~M.}\ \bibnamefont {Neal}},\ }\bibfield  {title} {\bibinfo {title} {{MCMC} using {Hamiltonian} dynamics},\ }in\ \href@noop {} {\emph {\bibinfo {booktitle} {Handbook of {Markov Chain Monte Carlo}}}},\ \bibinfo {editor} {edited by\ \bibinfo {editor} {\bibfnamefont {S.}~\bibnamefont {Brooks}}, \bibinfo {editor} {\bibfnamefont {A.}~\bibnamefont {Gelman}}, \bibinfo {editor} {\bibfnamefont {G.}~\bibnamefont {Jones}},\ and\ \bibinfo {editor} {\bibfnamefont {X.-L.}\ \bibnamefont {Meng}}}\ (\bibinfo  {publisher} {Chapman and Hall/CRC},\ \bibinfo {year} {2011})\ pp.\ \bibinfo {pages} {113--162}\BibitemShut {NoStop}%
\bibitem [{\citenamefont {Aziz}\ \emph {et~al.}(1979)\citenamefont {Aziz}, \citenamefont {Nain}, \citenamefont {Carley}, \citenamefont {Taylor},\ and\ \citenamefont {McConville}}]{aziz1979accurate}%
  \BibitemOpen
  \bibfield  {author} {\bibinfo {author} {\bibfnamefont {R.~A.}\ \bibnamefont {Aziz}}, \bibinfo {author} {\bibfnamefont {V.~P.~S.}\ \bibnamefont {Nain}}, \bibinfo {author} {\bibfnamefont {J.~S.}\ \bibnamefont {Carley}}, \bibinfo {author} {\bibfnamefont {W.~L.}\ \bibnamefont {Taylor}},\ and\ \bibinfo {author} {\bibfnamefont {G.~T.}\ \bibnamefont {McConville}},\ }\bibfield  {title} {\bibinfo {title} {An accurate intermolecular potential for helium},\ }\href@noop {} {\bibfield  {journal} {\bibinfo  {journal} {J. Chem. Phys.}\ }\textbf {\bibinfo {volume} {70}},\ \bibinfo {pages} {4330} (\bibinfo {year} {1979})}\BibitemShut {NoStop}%
\bibitem [{\citenamefont {Weeks}\ \emph {et~al.}(1971)\citenamefont {Weeks}, \citenamefont {Chandler},\ and\ \citenamefont {Andersen}}]{weeks1971role}%
  \BibitemOpen
  \bibfield  {author} {\bibinfo {author} {\bibfnamefont {J.~D.}\ \bibnamefont {Weeks}}, \bibinfo {author} {\bibfnamefont {D.}~\bibnamefont {Chandler}},\ and\ \bibinfo {author} {\bibfnamefont {H.~C.}\ \bibnamefont {Andersen}},\ }\bibfield  {title} {\bibinfo {title} {Role of repulsive forces in determining the equilibrium structure of simple liquids},\ }\href@noop {} {\bibfield  {journal} {\bibinfo  {journal} {J. Chem. Phys.}\ }\textbf {\bibinfo {volume} {54}},\ \bibinfo {pages} {5237} (\bibinfo {year} {1971})}\BibitemShut {NoStop}%
\bibitem [{\citenamefont {Hollingsworth}\ and\ \citenamefont {Dror}(2018)}]{hollingsworth2018molecular}%
  \BibitemOpen
  \bibfield  {author} {\bibinfo {author} {\bibfnamefont {S.~A.}\ \bibnamefont {Hollingsworth}}\ and\ \bibinfo {author} {\bibfnamefont {R.~O.}\ \bibnamefont {Dror}},\ }\bibfield  {title} {\bibinfo {title} {Molecular dynamics simulation for all},\ }\href@noop {} {\bibfield  {journal} {\bibinfo  {journal} {Neuron}\ }\textbf {\bibinfo {volume} {99}},\ \bibinfo {pages} {1129} (\bibinfo {year} {2018})}\BibitemShut {NoStop}%
\bibitem [{\citenamefont {Walker}(1974)}]{walker1974new}%
  \BibitemOpen
  \bibfield  {author} {\bibinfo {author} {\bibfnamefont {A.~J.}\ \bibnamefont {Walker}},\ }\bibfield  {title} {\bibinfo {title} {New fast method for generating discrete random numbers with arbitrary frequency distributions},\ }\href@noop {} {\bibfield  {journal} {\bibinfo  {journal} {Electron. Lett.}\ }\textbf {\bibinfo {volume} {10}},\ \bibinfo {pages} {127} (\bibinfo {year} {1974})}\BibitemShut {NoStop}%
\bibitem [{\citenamefont {Swendsen}\ and\ \citenamefont {Wang}(1986)}]{swendsen1986replica}%
  \BibitemOpen
  \bibfield  {author} {\bibinfo {author} {\bibfnamefont {R.~H.}\ \bibnamefont {Swendsen}}\ and\ \bibinfo {author} {\bibfnamefont {J.-S.}\ \bibnamefont {Wang}},\ }\bibfield  {title} {\bibinfo {title} {Replica {Monte Carlo} simulation of spin glasses},\ }\href@noop {} {\bibfield  {journal} {\bibinfo  {journal} {Phys. Rev. Lett.}\ }\textbf {\bibinfo {volume} {57}},\ \bibinfo {pages} {2607} (\bibinfo {year} {1986})}\BibitemShut {NoStop}%
\bibitem [{\citenamefont {Hansmann}(1997)}]{hansmann1997parallel}%
  \BibitemOpen
  \bibfield  {author} {\bibinfo {author} {\bibfnamefont {U.~H.~E.}\ \bibnamefont {Hansmann}},\ }\bibfield  {title} {\bibinfo {title} {Parallel tempering algorithm for conformational studies of biological molecules},\ }\href@noop {} {\bibfield  {journal} {\bibinfo  {journal} {Chem. Phys. Lett.}\ }\textbf {\bibinfo {volume} {281}},\ \bibinfo {pages} {140} (\bibinfo {year} {1997})}\BibitemShut {NoStop}%
\bibitem [{\citenamefont {Hukushima}(1999)}]{hukushima1999domain}%
  \BibitemOpen
  \bibfield  {author} {\bibinfo {author} {\bibfnamefont {K.}~\bibnamefont {Hukushima}},\ }\bibfield  {title} {\bibinfo {title} {Domain-wall free energy of spin-glass models: Numerical method and boundary conditions},\ }\href@noop {} {\bibfield  {journal} {\bibinfo  {journal} {Phys. Rev. E}\ }\textbf {\bibinfo {volume} {60}},\ \bibinfo {pages} {3606} (\bibinfo {year} {1999})}\BibitemShut {NoStop}%
\bibitem [{\citenamefont {Katzgraber}\ \emph {et~al.}(2006)\citenamefont {Katzgraber}, \citenamefont {Trebst}, \citenamefont {Huse},\ and\ \citenamefont {Troyer}}]{katzgraber2006feedback}%
  \BibitemOpen
  \bibfield  {author} {\bibinfo {author} {\bibfnamefont {H.~G.}\ \bibnamefont {Katzgraber}}, \bibinfo {author} {\bibfnamefont {S.}~\bibnamefont {Trebst}}, \bibinfo {author} {\bibfnamefont {D.~A.}\ \bibnamefont {Huse}},\ and\ \bibinfo {author} {\bibfnamefont {M.}~\bibnamefont {Troyer}},\ }\bibfield  {title} {\bibinfo {title} {Feedback-optimized parallel tempering {Monte Carlo}},\ }\href@noop {} {\bibfield  {journal} {\bibinfo  {journal} {J. Stat. Mech.}\ }\textbf {\bibinfo {volume} {2006}},\ \bibinfo {pages} {P03018} (\bibinfo {year} {2006})}\BibitemShut {NoStop}%
\bibitem [{\citenamefont {Sugita}\ \emph {et~al.}(2000)\citenamefont {Sugita}, \citenamefont {Kitao},\ and\ \citenamefont {Okamoto}}]{Sugita2000MultidimensionalREM}%
  \BibitemOpen
  \bibfield  {author} {\bibinfo {author} {\bibfnamefont {Y.}~\bibnamefont {Sugita}}, \bibinfo {author} {\bibfnamefont {A.}~\bibnamefont {Kitao}},\ and\ \bibinfo {author} {\bibfnamefont {Y.}~\bibnamefont {Okamoto}},\ }\bibfield  {title} {\bibinfo {title} {Multidimensional replica-exchange method for free-energy calculations},\ }\href {https://doi.org/10.1063/1.1308516} {\bibfield  {journal} {\bibinfo  {journal} {The Journal of Chemical Physics}\ }\textbf {\bibinfo {volume} {113}},\ \bibinfo {pages} {6042} (\bibinfo {year} {2000})}\BibitemShut {NoStop}%
\bibitem [{\citenamefont {Fukunishi}\ \emph {et~al.}(2002)\citenamefont {Fukunishi}, \citenamefont {Watanabe},\ and\ \citenamefont {Takada}}]{Fukunishi2002HamiltonianREM}%
  \BibitemOpen
  \bibfield  {author} {\bibinfo {author} {\bibfnamefont {H.}~\bibnamefont {Fukunishi}}, \bibinfo {author} {\bibfnamefont {O.}~\bibnamefont {Watanabe}},\ and\ \bibinfo {author} {\bibfnamefont {S.}~\bibnamefont {Takada}},\ }\bibfield  {title} {\bibinfo {title} {On the hamiltonian replica exchange method for efficient sampling of biomolecular systems: Application to protein structure prediction},\ }\href {https://doi.org/10.1063/1.1472510} {\bibfield  {journal} {\bibinfo  {journal} {The Journal of Chemical Physics}\ }\textbf {\bibinfo {volume} {116}},\ \bibinfo {pages} {9058} (\bibinfo {year} {2002})}\BibitemShut {NoStop}%
\bibitem [{\citenamefont {Ceperley}(1995)}]{ceperley1995path}%
  \BibitemOpen
  \bibfield  {author} {\bibinfo {author} {\bibfnamefont {D.~M.}\ \bibnamefont {Ceperley}},\ }\bibfield  {title} {\bibinfo {title} {Path integrals in the theory of condensed helium},\ }\href@noop {} {\bibfield  {journal} {\bibinfo  {journal} {Rev. Mod. Phys.}\ }\textbf {\bibinfo {volume} {67}},\ \bibinfo {pages} {279} (\bibinfo {year} {1995})}\BibitemShut {NoStop}%
\bibitem [{\citenamefont {Rathore}\ \emph {et~al.}(2005)\citenamefont {Rathore}, \citenamefont {Chopra},\ and\ \citenamefont {de~Pablo}}]{rathore2005optimal}%
  \BibitemOpen
  \bibfield  {author} {\bibinfo {author} {\bibfnamefont {N.}~\bibnamefont {Rathore}}, \bibinfo {author} {\bibfnamefont {M.}~\bibnamefont {Chopra}},\ and\ \bibinfo {author} {\bibfnamefont {J.~J.}\ \bibnamefont {de~Pablo}},\ }\bibfield  {title} {\bibinfo {title} {Optimal allocation of replicas in parallel tempering simulations},\ }\href@noop {} {\bibfield  {journal} {\bibinfo  {journal} {J. Chem. Phys.}\ }\textbf {\bibinfo {volume} {122}},\ \bibinfo {pages} {024111} (\bibinfo {year} {2005})}\BibitemShut {NoStop}%
\bibitem [{\citenamefont {Hamze}\ \emph {et~al.}(2010)\citenamefont {Hamze}, \citenamefont {Dickson},\ and\ \citenamefont {Karimi}}]{hamze2010robust}%
  \BibitemOpen
  \bibfield  {author} {\bibinfo {author} {\bibfnamefont {F.}~\bibnamefont {Hamze}}, \bibinfo {author} {\bibfnamefont {N.}~\bibnamefont {Dickson}},\ and\ \bibinfo {author} {\bibfnamefont {K.}~\bibnamefont {Karimi}},\ }\bibfield  {title} {\bibinfo {title} {Robust parameter selection for parallel tempering},\ }\href@noop {} {\bibfield  {journal} {\bibinfo  {journal} {Int. J. Mod. Phys. C}\ }\textbf {\bibinfo {volume} {21}},\ \bibinfo {pages} {603} (\bibinfo {year} {2010})}\BibitemShut {NoStop}%
\bibitem [{\citenamefont {Krauth}(2006)}]{krauth2006statistical}%
  \BibitemOpen
  \bibfield  {author} {\bibinfo {author} {\bibfnamefont {W.}~\bibnamefont {Krauth}},\ }\href@noop {} {\emph {\bibinfo {title} {Statistical Mechanics: Algorithms and Computations}}}\ (\bibinfo  {publisher} {Oxford University Press},\ \bibinfo {year} {2006})\BibitemShut {NoStop}%
\bibitem [{\citenamefont {Landau}\ \emph {et~al.}(2004)\citenamefont {Landau}, \citenamefont {Tsai},\ and\ \citenamefont {Exler}}]{landau2004new}%
  \BibitemOpen
  \bibfield  {author} {\bibinfo {author} {\bibfnamefont {D.~P.}\ \bibnamefont {Landau}}, \bibinfo {author} {\bibfnamefont {S.-H.}\ \bibnamefont {Tsai}},\ and\ \bibinfo {author} {\bibfnamefont {M.}~\bibnamefont {Exler}},\ }\bibfield  {title} {\bibinfo {title} {A new approach to {Monte Carlo} simulations in statistical physics: {W}ang--{L}andau sampling},\ }\href@noop {} {\bibfield  {journal} {\bibinfo  {journal} {Am. J. Phys.}\ }\textbf {\bibinfo {volume} {72}},\ \bibinfo {pages} {1294} (\bibinfo {year} {2004})}\BibitemShut {NoStop}%
\bibitem [{\citenamefont {Ambegaokar}\ and\ \citenamefont {Troyer}(2010)}]{ambegaokar2010estimating}%
  \BibitemOpen
  \bibfield  {author} {\bibinfo {author} {\bibfnamefont {V.}~\bibnamefont {Ambegaokar}}\ and\ \bibinfo {author} {\bibfnamefont {M.}~\bibnamefont {Troyer}},\ }\bibfield  {title} {\bibinfo {title} {Estimating errors reliably in {Monte Carlo} simulations of the {Ehrenfest} model},\ }\href@noop {} {\bibfield  {journal} {\bibinfo  {journal} {Am. J. Phys.}\ }\textbf {\bibinfo {volume} {78}},\ \bibinfo {pages} {150} (\bibinfo {year} {2010})}\BibitemShut {NoStop}%
\bibitem [{\citenamefont {Harada}(2011)}]{Harada2011}%
  \BibitemOpen
  \bibfield  {author} {\bibinfo {author} {\bibfnamefont {K.}~\bibnamefont {Harada}},\ }\bibfield  {title} {\bibinfo {title} {Bayesian inference in the scaling analysis of critical phenomena},\ }\href@noop {} {\bibfield  {journal} {\bibinfo  {journal} {Phys. Rev. E}\ }\textbf {\bibinfo {volume} {84}},\ \bibinfo {pages} {056704} (\bibinfo {year} {2011})}\BibitemShut {NoStop}%
\bibitem [{\citenamefont {Sindzingre}\ \emph {et~al.}(1989{\natexlab{b}})\citenamefont {Sindzingre}, \citenamefont {Klein},\ and\ \citenamefont {Ceperley}}]{Sindzingre1989HeClusters}%
  \BibitemOpen
  \bibfield  {author} {\bibinfo {author} {\bibfnamefont {P.}~\bibnamefont {Sindzingre}}, \bibinfo {author} {\bibfnamefont {M.~L.}\ \bibnamefont {Klein}},\ and\ \bibinfo {author} {\bibfnamefont {D.~M.}\ \bibnamefont {Ceperley}},\ }\bibfield  {title} {\bibinfo {title} {Path-integral monte carlo study of low-temperature \({}^{4}\mathrm{He}\) clusters},\ }\href {https://doi.org/10.1103/PhysRevLett.63.1601} {\bibfield  {journal} {\bibinfo  {journal} {Physical Review Letters}\ }\textbf {\bibinfo {volume} {63}},\ \bibinfo {pages} {1601} (\bibinfo {year} {1989}{\natexlab{b}})}\BibitemShut {NoStop}%
\bibitem [{\citenamefont {Gordillo}\ and\ \citenamefont {Ceperley}(1998)}]{Gordillo1998HeliumFilm}%
  \BibitemOpen
  \bibfield  {author} {\bibinfo {author} {\bibfnamefont {M.~C.}\ \bibnamefont {Gordillo}}\ and\ \bibinfo {author} {\bibfnamefont {D.~M.}\ \bibnamefont {Ceperley}},\ }\bibfield  {title} {\bibinfo {title} {Path-integral calculation of the two-dimensional phase diagram of \({}^{4}\mathrm{He}\)},\ }\href {https://doi.org/10.1103/PhysRevB.58.6447} {\bibfield  {journal} {\bibinfo  {journal} {Physical Review B}\ }\textbf {\bibinfo {volume} {58}},\ \bibinfo {pages} {6447} (\bibinfo {year} {1998})}\BibitemShut {NoStop}%
\end{thebibliography}
\end{document}